\documentclass[12pt]{article}
\usepackage{epsf,amssymb,psfrag}

\catcode`\@=11
\def \thesection {\arabic{section}.}

\def \be  {\begin{equation}}
\def \ee  {\end{equation}}
\def \ba  {\begin{eqnarray}}
\def \ea  {\end{eqnarray}}
\def \baa {\begin{eqnarray*}}
\def \eaa {\end{eqnarray*}}
\def \bb  {\begin {thebibliography} }
\def \eb  {\end{thebibliography}}
\def \lab #1 {\label{#1}}

\newcommand\re[1]{({\ref{#1}})}

\def \qqqquad {\qquad\qquad}
\def \matrix #1 {\left(\begin{array}{cc} #1 \end{array}\right)}

\def \tr {\mathop{\rm tr}\nolimits}
\def \Im {\mathop{\rm Im}\nolimits}
\def \Re {\mathop{\rm Re}\nolimits}

\def \e  {\mathop{\rm e}\nolimits}
\newcommand\lr[1]{{\left({#1}\right)}}

\newcommand \vev [1] {\langle{#1}\rangle}

\newcommand \ket [1] {|{#1}\rangle}

\newcommand{\as}{\ifmmode\alpha_{\rm s}\else{$\alpha_{\rm s}$}\fi}
\newcommand{\asbar}{\ifmmode\bar{\alpha}_{\rm s}\else{$\bar{\alpha}_{\rm s}$}\fi}

\def \CH {{\cal H}}

\def \CV {{\cal V}}

\def \CD {{\cal D}}

\font\cmss=cmss12 
\def\inbar{\,\vrule height1.5ex width.4pt depth0pt}
\def\IC{\relax\hbox{$\inbar\kern-.3em{\rm C}$}}
\def\IZ{\relax{\hbox{\cmss Z\kern-.4em Z}}}
\def\IR{{\hbox{{\rm I}\kern-.2em\hbox{\rm R}}}}

\def\IP{{\hbox{{\rm I}\kern-.2em\hbox{\rm P}}}}
\def\II{\hbox{{1}\kern-.25em\hbox{l}}}

\def\numberbysection{\@addtoreset{equation}{section}
                     \def\theequation{\thesection\arabic{equation}}}
\numberbysection

\newcommand \mybf[1] {\mbox{\boldmath$\scriptstyle {#1} $}}
\newcommand \Mybf[1] {\mbox{\boldmath$ {#1} $}}
\newbox\lett\newdimen\lheight\newdimen\lwidth
\def\ontop#1#2{\setbox\lett=\hbox{#2}\lheight\ht\lett
\multiply\lheight by 12 \divide\lheight by 10\relax%
\lwidth\wd\lett \multiply\lwidth by 8 \divide\lwidth by 10\relax #2\kern-\lwidth%
\raise\lheight\hbox{{$\scriptstyle #1$}}\kern.1ex}

\begin{document}

\begin{titlepage}
\begin{flushright}
\begin{tabular}{l}
LPT--Orsay--03--61\\
RUB-TP2-12/03\\
hep-th/0309144
\end{tabular}
\end{flushright}

\vskip3cm
\begin{center}
  {\large \bf Baxter $\mathbb{Q}-$operator and Separation of Variables \\[3mm]
for the open $SL(2,\mathbb{R})$ spin chain }

\def\thefootnote{\fnsymbol{footnote}}%
\vspace{1cm}
{\sc S.\'{E}. Derkachov}${}^1$, {\sc G.P.~Korchemsky}${}^2$
 and {\sc A.N.~Manashov}${}^3$\footnote{ Permanent
address:\ Department of Theoretical Physics,  Sankt-Petersburg State University,
St.-Petersburg, Russia}
\\[0.5cm]

\vspace*{0.1cm} ${}^1$ {\it
Department of Mathematics, St.-Petersburg Technology Institute,\\
St.-Petersburg, Russia
                       } \\[0.2cm]
\vspace*{0.1cm} ${}^2$ {\it
Laboratoire de Physique Th\'eorique%
\footnote{Unite Mixte de Recherche du CNRS (UMR 8627)},
Universit\'e de Paris XI, \\
91405 Orsay C\'edex, France
                       } \\[0.2cm]
\vspace*{0.1cm} ${}^3$
 {\it
Institut f\"ur Theoretische Physik II, Ruhr-Universit\"at Bochum,\\
D-44780 Bochum, Germany}

\vskip2cm

{\bf Abstract:\\[10pt]} \parbox[t]{\textwidth}{
We construct the Baxter $\mathbb{Q}-$operator and the representation of the
Separated Variables (SoV) for the homogeneous open $SL(2,\mathbb{R})$ spin chain.
Applying the diagrammatical approach, we calculate Sklyanin's integration measure
in the separated variables and obtain the solution to the spectral problem for
the model in terms of the eigenvalues of the $\mathbb{Q}-$operator. We show that
the transition kernel to the SoV representation is factorized into the product of
certain operators each depending on a single separated variable. As a
consequence, it has a universal pyramid-like form that has been already observed
for various quantum integrable models such as periodic Toda chain, closed
$SL(2,\mathbb{R})$ and $SL(2,\mathbb{C})$ spin chains.} \vskip1cm

\end{center}

\end{titlepage}

\newpage
\tableofcontents

\setcounter{footnote}{0}

\section{Introduction}

Recently it has been found that the evolution equations describing the scale
dependence of certain correlation functions in four-dimensional Yang-Mills theory
possess a hidden symmetry. Remarkably enough, the emerging integrable structures
are well-known in the theory of lattice integrable models \cite{QISM} as
corresponding to open Heisenberg spin magnets. In particular, the energy spectrum
of the magnet determines the spectrum of the anomalous dimensions of the
correlation functions in Yang-Mills theory~\cite{BDM,AB,DKM99}. A unusual feature
of these models as compared with conventional magnets studied thoroughly in
applications to statistical physics~\cite{Baxter} is that the spin operators are
generators of infinite-dimensional representations of the $SL(2,\mathbb{R})$
group. This group emerges as subgroup of the full $SO(4,2)$ conformal group of
four-dimensional Yang-Mills theory.

Exact solution of the spectral problem for integrable systems with
infinite-dimensional quantum space is a nontrivial task. The conventional
Algebraic Bethe Ansatz (ABA)~\cite{ABA} is not always applicable to such systems
and one has to rely instead on a more elaborated methods like the Baxter
$\mathbb{Q}-$operator~\cite{Baxter} and the Separation of Variables
(SoV)~\cite{Sklyanin}. Being combined together, the two methods allow one to find
the energy spectrum of the model and obtain integral representation for the
eigenstates. At present, such program has been carried out for a number of models
with periodic boundary conditions. They include periodic Toda chain~\cite{PG,KL},
the DST model~\cite{KSS}, noncompact closed $SL(2)$ Heisenberg
magnets~\cite{SD,DKM,SoV} and Calogero-Sutherland model~\cite{KMS}. In the
present paper, we apply the both methods to the quantum $SL(2,\mathbb{R})$ open
Heisenberg spin chain.

A systematic approach to building quantum integrable models with nontrivial
boundary conditions (including open Heisenberg spin chains)
has been developed by Sklyanin~\cite{Sklyanin88}. For such models a little
progress has been made in constructing the $\mathbb{Q}-$operator and the SoV
representation. One of the reasons for this is that the $R-$matrix formulation is
more cumbersome in that case as compared to the models with periodic boundary
conditions and, in addition, there exist no regular procedure for obtaining the
$\mathbb{Q}-$operator.

In this paper, we construct the Baxter $\mathbb{Q}-$operator and representation
of the Separated variables for the quantum $SL(2,\mathbb{R})$ open spin chain.
Our analysis is based on the Feynman diagram approach described at length in
previous publications~\cite{DKM,SoV}. In this approach, one realizes the
$\mathbb{Q}-$operator as an integral operator acting on the quantum space of the
model and represents its kernel as a certain Feynman diagram. Then, various
properties of the $\mathbb{Q}-$operator can be established by making use of a few
elementary diagrammatical relations. Using the obtained expressions, we determine
the energy spectrum of the open Heisenberg spin chain in terms of the eigenvalues
of the $\mathbb{Q}-$operator and obtain integral representation for the
eigenfunctions.

The presentation is organized as follows. In Section~2 we define the open
Heisenberg magnet with the $SL(2,\mathbb{R})$ spin symmetry and review Sklyanin's
formulation of the model. In Section~3 we construct Baxter $\mathbb{Q}-$operator
for the homogeneous open spin chain and establish its properties. In Section~4 we
present an explicit construction of the unitary transformation to the Separated
Variables for the open $SL(2,\mathbb{R})$ spin chain. In particular, we calculate
the integration measure defining the scalar product in the SoV representation and
discuss its
analytical properties. 
In Section~5 we demonstrate that for the open spin chain with two sites the
eigenvalues of the $\mathbb{Q}-$operator coincide with the Wilson orthogonal
polynomials. Section~6 contains concluding remarks. Some technical details and
description of the diagrammatical technique are given in the Appendix.

\setcounter{equation}{0}
\section{Open Heisenberg spin chain}
\label{Open}

\subsection{Definition of the model}

The homogeneous open Heisenberg spin chain is a lattice model of $N$ interacting
spins $\vec S_n=(S_n^1,S_n^2,S_n^3)$ (with $n=1,...,N$) described by the
Hamiltonian
\be\label{H}
\CH_N = \sum_{n=1}^{N-1} H_{n,n+1}, \qqqquad
H_{n,n+1}~=~2\left[\psi(J_{n,n+1})-\psi(2s)\right]\,,
\ee
where $\psi(x)=d\log\Gamma(z)/dz$ is the Euler $\psi-$function. The pairwise
Hamiltonian $H_{n,n+1}$ defines the interaction between two neighboring spins
$\vec S_n$ and $\vec S_{n+1}$. It is expressed in terms of the operator
$J_{n,n+1}$ related to their sum
\be\label{JJ}
J_{n,n+1}(J_{n,n+1}-1)=(\vec{S}_n+\vec{S}_{n+1})^2\,.
\ee
The spin operators in different sites commute with each other and obey the
standard commutation relations
\be
[S_n^a,S_k^b]=i\varepsilon_{abc}\delta_{nk} S_n ^c\,,\qquad \vec S^2_n
= s_n(s_n-1)\,.
\ee
We shall assume for simplicity that the spin chain is homogeneous,
$s_1=...=s_N=s$, with real $s\ge 1/2$ the same as in \re{H}.

Notice that the Hamiltonian \re{H} does not involve interaction between the
boundary spins $\vec S_1$ and $\vec S_N$. If one added the corresponding
two-particle Hamiltonian $H_{N,1}$ to the r.h.s.\ of \re{H}, the resulting
Hamiltonian would define a homogeneous closed Heisenberg spin chain. The latter
model admits solution within the $R-$matrix approach both by the ABA
method~\cite{ABA} and the methods of the Baxter $\mathbb{Q}-$operator~\cite{SD}
and SoV~\cite{SoV}. In the present paper we extend the analysis performed in the
papers~\cite{SD,SoV} to the case of the open spin chain and apply the method of
the Baxter $\mathbb{Q}-$operator to solve
the spectral problem for the Hamiltonian \re{H}
\be
\mathcal{H}_N \Psi_{\mybf{q}}(z_1,\ldots,z_N) =
E_{\mybf{q}}\Psi_{\mybf{q}}(z_1,\ldots,z_N)\,.
\label{Sch}
\ee
Here $\Mybf{q}$ denotes the complete set of the quantum numbers parameterizing
the energy spectrum  and $z_n$ (with $n=1,\ldots,N$) are the coordinates on the
quantum space $V_n$ associated with the $n$th site of the spin chain.

The Hamiltonian $\mathcal{H}_N$ acts on the quantum space of the model
$\mathcal{V}_N=\prod_{n=1}^N \otimes V_n$
and its energy spectrum depends on the choice of the Hilbert space $V_n$.
In what follows we shall assume that
$\mathcal{V}_N$ is spanned by functions $\Psi(z_1,\ldots,z_N)\in \mathcal{V}_N$
holomorphic in the upper half-plane $\Im z_n>0$ and normalizable with respect to
the scalar product
\begin{equation}
\vev{\Psi_1|\Psi_2}~=~\int\mathcal{D}^N z\,
\left(\Psi_1(z_1,\ldots,z_N)\right)^*\,\Psi_2(z_1,\ldots,z_N)\,,
\label{norm}
\end{equation} where integration measure is defined as $\mathcal{D}^N
z=\prod_{n=1}^N \mathcal{D} z_n$ with ($z_n=x_n+iy_n$)
\be
\mathcal{D}z_n
=\frac{2s-1}{\pi}\,d^2z_n \, (2\Im z_n)^{2s-2}\theta(\Im z_n)
=\frac{2s-1}{\pi}\,dx_ndy_n\,(2y_n)^{2s-2}\theta(y_n)
\label{measure}
\ee
and integration in \re{norm} goes over the upper half-plane. The spin operators
$\vec S_n$ can be realized on this space as differential operators%
\footnote{In Yang-Mills theory the spin operators \re{s-rep} define
representation of the generators of the collinear $SL(2,\mathbb{R})$ subgroup of
the full $SO(4,2)$ conformal group on the space of correlation functions
$\vev{0|\Phi_s(z_1 n)\ldots \Phi_s(z_N n)|0}$ of primary fields with conformal
spin $s$ and ``living'' on the light-cone $n_\mu^2=0$.}
\begin{equation}
\label{s-rep}
S_n^{+}~=~z_n^2\partial_{z_n}+2 s\,z_n,\ \ \
S_n^{-}~=~-\partial_{z_n},\ \ \
S_n^{0}~=~z_n\partial_{z_n}+ s\,.
\end{equation}
where $S_n^\pm =S_n^1\pm iS_n^2$ and $S_n^0=S_n^3$. These operators are
anti-hermitian with respect to the scalar product \re{norm}
\be
(S_n^0)^\dagger = -S_n^0\,,\qquad(S_n^\pm)^\dagger = -S_n^\pm\,.
\ee
Notice that the quantum space of the model is
infinite-dimensional for arbitrary finite $N$. For integer and half-integer $s$,
the Hilbert space $V_n$ coincides with the representation space of unitary
representation of the $SL(2,\mathbb{R})$ group of the discrete series~\cite{Gelfand}.

\subsection{$R-$matrix formulation}

The open $SL(2,\mathbb{R})$ Heisenberg spin magnet \re{H} is a completely
integrable model. To identify its integrals of motion we follow Sklyanin's
approach~\cite{Sklyanin88}. To begin with, one defines the Lax operator for the
$SL(2,\mathbb{R})$ magnet
\begin{equation}
\label{Lax}
L_n(u)=u+i (\vec\sigma\cdot  \vec{S}_n)= \left(\begin{array}{cc}u+iS^0_n&
iS^-_n\\iS^+_n&u-iS^0_n\end{array}\right),
\end{equation}
where $\vec \sigma=(\sigma_1,\sigma_2,\sigma_3)$ are the Pauli matrices.
 It acts on the tensor product of the auxiliary space and the quantum space
in the $n$th site, ${\mathbb C}^2\otimes{V_n}$. Taking the product of $N$ Lax
operators along the spin chain in the auxiliary space one defines the operator
-- the monodromy matrix for the closed spin chain
\be\label{Tc}
T_N(u)~=~L_1(u)\ldots
L_N(u)~=~\left(\begin{array}{cc}a(u)&b(u)\\
c(u)&d(u)
\end{array}\right),
\ee
which is a $2\times 2$ matrix with the entries $a(u),\ldots,d(u)$ being operators
acting on $\mathcal{V}_N$. It satisfies the Yang-Baxter commutation relations
\be\label{YBR}
R_{12}(u-v)\ontop1T_N(u)\ontop2T_N(v)=\ontop2T_N(v)\ontop1T_N(u)R_{12}(u-v)\,,
\ee
where  $\ontop1T_N(u)={T}_N(u)\otimes \II$ and $\ontop2T_N(v)=\II\otimes
{T}_N(v)$. The $R-$matrix acts on the tensor product of two auxiliary spaces,
$\mathbb{C}^2\otimes\mathbb{C}^2$,
\be\label{Rm}
R_{12}(u) = u\,\II + i\,P_{12}\,,
\ee
with $P_{12}$ being the permutation operator. The monodromy matrix for the open
spin chain is defined as~\cite{Sklyanin88} \footnote{General definition of the
integrable spin chain with nontrivial boundary conditions involves the boundary
matrices $K_\pm$~\cite{Ch,Sklyanin88,KS}. The Hamiltonian \re{H} corresponds to
the simplest case $K_\pm = \II$.}
\be\label{To}
\mathbb{T}_N(u)~=~T_N(u)\,T_N^{-1}(-u+i)=\frac1{\rho^{N}(u)}\cdot
T_N(u)\,\sigma_2\,T^{t}_N(-u)\,\sigma_2\,,
\ee
where the c-valued factor $\rho(u)=(u-is)(u+i(s-1))$ absorbs all poles of
$\mathbb{T}_N(u)$ and the superscript `$t$' denotes transposition in the
auxiliary space. It satisfies the fundamental ``reflection'' Yang-Baxter
relation~\cite{Ch,Sklyanin88,KS}
\be\label{YBo}
\mathbb{\ontop2T}_N(v)\,R_{12}(u+v-i)\,\mathbb{\ontop1T}_N(u)\,R_{12}(u-v)~=~
R_{12}(u-v)\,\mathbb{\ontop1T}_N(u)\,R_{12}(u+v-i)\,\mathbb{\ontop2T}_N(v)\,
\ee
with the same $R-$matrix \re{Rm}. It proves convenient to change a normalization
of $\mathbb{T}_N(u)$ as
\be\label{nTo}
\widehat{\mathbb{T}}_N(u)=\rho^N(-u)\,\mathbb{T}_N(-u)=T_N(-u)\,
\sigma_2\,T^t_N(u)\,\sigma_2=
\left(\begin{array}{cc}A(u)&B(u)\\
C(u)&D(u)
\end{array}\right).
\ee
It follows from \re{To} that $\widehat{\mathbb{T}}_N(u)$ satisfies the relation
\be
\widehat{\mathbb{T}}_N(-u-i)\,\widehat{\mathbb{T}}_N(u)=\left[(u+is)(u-i(s-1))\right]^{2N}
\II\,.
\label{q-det}
\ee
The Yang-Baxter relation \re{YBo} leads to the set of fundamental relations for
the operators $A(u),\ldots,D(u)$. For our purposes we will need only two of them
\ba
&& B(u)B(v) = B(v)B(u)\,,
\label{BB}
\\
&& B(u)D(v)= \frac{(u+v+i)(u-v-i)}{(u-v)(u+v)}D(v)B(u) + i
\left[A(u)+\frac{u+v+i}{u-v}D(u)\right] \frac{B(v)}{u+v}\,. \nonumber
\ea
The monodromy matrix \re{nTo} satisfies the following relation
\be
\widehat{\mathbb{T}}_N(u)=\frac{1}{2u-i}\left[ 2u\,\sigma_2
\widehat{\mathbb{T}}_N^{\,t}(-u)\sigma_2-i\, \widehat{\mathbb{T}}_N(-u)\right].
\label{T-rel}
\ee
To verify it one starts with the definition of $\widehat{\mathbb{T}}_N(u)$,
Eq.~\re{nTo}, interchanges the operators $T_N(-u)$ and $T^t_N(u)$ with a help of
the Yang-Baxter relation \re{YBR} and uses the explicit expression \re{Rm} for
the $R-$matrix. Substitution of \re{nTo} into \re{T-rel} yields
\be
D(u)=\frac{1}{2u-i}\left[2u A(-u)-iD(-u)\right]\,,\qquad
\frac{B(u)}{2u+i}=\frac{B(-u)}{-2u+i}\,.
\label{A-minus}
\ee
In the standard manner, the transfer matrix for the open spin chain
$\widehat{t}_N(u)$ is defined as the trace of the monodromy matrix \re{nTo} over
the auxiliary space
\ba\label{th}
\widehat t_N(u) = \tr\widehat{\mathbb{T}}_N(u)=A(u) + D(u)
=\left(1-\frac{i}{2u}\right)\,D(u)~+~\left(1+\frac{i}{2u}\right)\,D(-u)\,,
\ea
where in the last relation we took into account \re{A-minus}. Following
Sklyanin~\cite{Sklyanin88} and making use of the Yang-Baxter relation \re{YBo},
one can show that the transfer matrix commutes with itself for different values
of the spectral parameter, with the Hamiltonian~(\ref{H}) and with the operator
of the total spin $\vec{S}=\sum_{n=1}^N \vec S_n$
\be\label{comm-th}
[\widehat t_N(u),\widehat t_N(v)]=[\widehat t_N(u),\CH_N]=[\widehat
t_N(u),\vec{S}]=0\,.
\ee
The expansion of $\widehat t_N(u)$ in powers of $u$ generates the integrals of
motion of the model. One deduces from \re{th} and \re{nTo} that the transfer
matrix is an even polynomial in $u$ of degree $2N$, $\widehat t_N(-u)=\widehat
t_N(u)$, which scales at large $u$ as $\widehat t_N(u)\sim 2(-1)^N u^{2N}$. In
addition, it follows from \re{To} that $\mathbb{T}_N(i/2)=\II$ leading to
\be
\widehat t_N(-i/2)=2\rho^N(i/2)=2(s-1/2)^{2N}\,.
\ee
These properties imply that $t_N(u)-t_N(\pm i/2)$ is proportional to
$(u+i/2)(u-i/2)$\\[2mm]
\be
\widehat t_N(u) = (-1)^N\left(u^2+1/4\right) \left[ 2 u^{2N-2} +\widehat q_2\,
u^{2N-4} + \ldots + \widehat q_{N-1} \, u^2 +\widehat q_N\right]+
2\left(s-1/2\right)^{2N}\,.
\label{q's}
\ee\\[2mm]
Here the $\widehat q-$operators are given by polynomials in the spin operators
$\vec S_n$, for instance
\be
\widehat q_2=-4 \vec S^2 + 2Ns(s-1)-\frac12\,.
\label{q2}
\ee
It follows from \re{comm-th} and \re{q's} that $N-1$ operators $\widehat
q_2,\ldots,\widehat q_N$ form the family of mutually commuting $SL(2)$ invariant
integrals of motion. Since $[\CH_N,\vec{S}]=0$, the remaining $N$th integral of
motion is provided by one of the components of the total spin. It is convenient
to choose the latter as $iS_-= -i\sum_n \partial_{z_n}$ since its eigenvalues
define the total momentum.

Thus, the open Heisenberg spin chain is a completely integrable model and the
spectral problem for the Hamiltonian~(\ref{H}) can be reformulated as the
spectral problem for the transfer matrix
\ba
\label{eit}
&&\widehat t_N(u)
\Psi_{\mybf{q},p}(z_1,\ldots,z_N)=t_N(u)\Psi_{\mybf{q},p}(z_1,\ldots,z_N)\,,
\\[2mm]
&& (iS_--p)\Psi_{\mybf{q},p}(z_1,\ldots,z_N)=0\,, \nonumber
\ea
where $t_N(u)$ is the eigenvalue of the transfer matrix~(\ref{th}) and
$\Mybf{q}=(q_2,\ldots,q_N)$ denotes the eigenvalues of the integrals of motion. A
general solution to \re{eit} takes the form
\be
\Psi_{\mybf{q},p}(z_1,\ldots,z_N)=\int_{-\infty}^\infty dx_0\, \e^{ipx_0}
\Psi_{\mybf{q}}(z_1-x_0,\ldots,z_N-x_0)\,,
\ee
where integration goes along the real axis. The eigenstate
$\Psi_{\mybf{q}}(z_1,\ldots,z_N)$ has to diagonalize simultaneously the operators
$\widehat q_2,\ldots,\widehat q_N$.

\setcounter{equation}{0}
\section{Baxter $\mathbb{Q}-$operator}
\label{QBaxter}

To solve the spectral problem for the open Heisenberg spin chain, Eq.~\re{eit},
we apply the method of the Baxter $\mathbb{Q}-$operator. The method relies on the
existence of the operator $\mathbb{Q}(u)$ which acts on the quantum space of the
model $\CV_N$, depends on the spectral parameter $u$ and satisfies the following
defining relations:
\begin{itemize}
\item Commutativity:
\be\label{Q-comm}
\left[\mathbb{Q}(u),\mathbb{Q}(v)\right]~=~0\,.
\ee
\item Q -- t relation:
\be\label{tQ}
\left[\mathbb{Q}(u),\widehat t_N(u)\right]=0\,.
\ee
\item Baxter relation:
\be\label{Bax-eq}
\widehat t_N(u)\,\mathbb{Q}(u)~=~\Delta_{+}(u)\,\mathbb{Q}(u+i)~+~
\Delta_{-}(u)\,\mathbb{Q}(u-i)\,,
\ee
\end{itemize}
where $\Delta_\pm(u)$ are some scalar functions of $u$. For the homogeneous open
spin chain they are given by
\be\label{Deltapm}
\Delta_\pm(u)=(-1)^N \frac{2u\mp i}{2u}(u\pm is)^{2N}\,.
\ee
In this Section, we construct the operator $\mathbb{Q}(u)$ satisfying
Eqs.~\re{Q-comm}--\re{Bax-eq} and discuss its properties.

It follows from \re{Q-comm} and \re{tQ} that the Baxter $\mathbb{Q}-$operator and
the transfer matrix $\widehat t_N(u)$ share the common set of the eigenstates
\be
\mathbb{Q}(u)\Psi_{\mybf{q}}(z_1,\ldots,z_N) = Q_{\mybf{q}}(u)
\Psi_{\mybf{q}}(z_1,\ldots,z_N)\,.
\label{Bax-eig}
\ee
The eigenstates $\Psi_{\mybf{q}}(z_1,\ldots,z_N)$ are the solutions to the
Schr\"odinger equation \re{Sch} whereas the corresponding eigenvalues of the
$\mathbb{Q}-$operator, $Q_{\mybf{q}}(u)$, satisfy the Baxter relation \re{Bax-eq}
with the transfer matrix $\widehat t_N(u)$, Eq.~\re{q's}, replaced by its
eigenvalue. As we will show below, the Baxter $\mathbb{Q}-$operator encodes
information about the spectrum of the open spin chain. Namely,
having calculated its eigenvalues $Q_{\mybf{q}}(u)$ one would be able to
reconstruct the energy spectrum of the model $E_{\mybf{q}}$.

\subsection{Gauge transformations}
\label{GT}

Our approach to constructing the Baxter $\mathbb{Q}-$operator is based on the
representation of $\mathbb{Q}(u)$ as an integral operator acting on the quantum
space of the model
\be\label{Bax-int}
\left[ \mathbb{Q}(u)\Psi\right](z_1,\ldots,z_N) ~=~\int \CD^N  w\,
Q_u(z_1\ldots,z_N|\bar w_1,\ldots,\bar w_N)\Psi(w_1,\ldots,w_N)\,,
\ee
with $\bar w_n=w_n^*$ and the integration measure defined in \re{measure}. To
find the explicit expression for the kernel $Q_u(z_1\ldots,z_N|\bar
w_1,\ldots,\bar w_N)$ we shall explore the fact that the transfer matrix of the
open spin chain, Eq.~\re{th}, is invariant under local gauge transformations of
the Lax operators~\cite{PG,SD}
\be\label{Lgr}
L_n(u)\to \widetilde L_n(u) = M_n^{-1}\,L_n(u)\,M_{n+1}\,,
\ee
where $M_n$ (with $n=1,\ldots,N+1$) are arbitrary $2\times 2$ matrices with $\det
M_n\neq 0$. According to Eqs.~\re{Tc}, \re{To} and \re{nTo} the operators
$T_N(u)$ and $\widehat{\mathbb{T}}_N(u)$ are transformed under \re{Lgr} as
\ba
T_N(u)&\to& \widetilde T_N(u)=M_1^{-1}\, T_N(u)\, M_{N+1},\ \ \ \ \ \ \
\nonumber
\\
\widehat{ \mathbb{T}}_N(u)&\to& \widetilde{\widehat{\mathbb{T}}}_N(u)=M_1^{-1}\,
\widehat{ \mathbb{T}}_N(u)\, M_{1}\,,
\label{Mo}
\ea
so that the transfer matrix $\widehat t_N(u)=\tr\widehat{\mathbb{T}}_N(u)$ stays
invariant.

The gauge rotation of the Lax operator, Eq.~\re{Lgr}, has been used by Pasquier
and Gaudin to construct the Baxter operator for the periodic Toda chain~\cite{PG}
and it was later applied to the closed spin chain in Refs.~\cite{SD,SoV}. In the
latter case, the transfer matrix equals $\tr T_N(u)$ and in order to preserve its
invariance under the transformation \re{Mo} one had to impose periodic boundary
conditions $M_{N+1}=M_1$. For the open spin chain the matrices $M_{N+1}$ and
$M_1$ can be arbitrary, $M_{N+1}\neq M_1$. \footnote{Notice that the monodromy
matrices $\widetilde T_N(u)$ and $\widetilde{\widehat{\mathbb{T}}}_N(u)$ satisfy
the Yang-Baxter relations, Eq~\re{YBR} and \re{YBo}, respectively. This follows
immediately from the invariance of the $R-$matrix~(\ref{Rm}) under
transformations $R\to U\, R\, U^{-1}$ with $U=M\otimes M$.} In spite of this
difference, many results obtained in Refs.~\cite{SD,SoV} for the closed spin
chain are applicable to the open chain.

To begin with, let us introduce the function~\cite{SD,SoV}
\be\label{Y-N}
Y_u(z_1,\ldots,z_N|\bar w_1,\ldots,\bar w_{N+1})~=~ \prod_{k=1}^N (z_k-\bar
w_k)^{-s-iu}\,(z_k-\bar w_{k+1})^{-s+iu}\,,
\ee
which is a (anti)holomorphic function of the complex variables $\vec
z=(z_1,\ldots,z_N)$ and $\vec w=(\bar w_1,\ldots,\bar w_{N+1})$  in the upper
half-plane $\Im z_k>0$ and $\Im w_n>0$ (with $\bar w_n=w_n^*$). It satisfies the
following relations
\ba
\widetilde b(u; \bar w_1,\bar w_{N+1})\,Y_u(\vec{z}|\vec{w})&=&0\,,\nonumber \\
\widetilde a(u; \bar w_1,\bar
w_{N+1})\,Y_u(\vec{z}|\vec{w})&=&(u+is)^N\,Y_{u+i}(\vec{z}|\vec{w})\,,
\label{TYc}
\\
\widetilde d(u; \bar w_1,\bar
w_{N+1})\,Y_u(\vec{z}|\vec{w})&=&(u-is)^N\,Y_{u-i}(\vec{z}|\vec{w})\,,\nonumber
\ea
where the operators $\widetilde a(u),\ldots,\widetilde d(u)$ are defined
similarly to \re{Tc} as the entries of the gauge rotated transfer matrix
$\widetilde T_N(u)$, Eq.~\re{Mo}, with the $M-$matrices given by
\be\label{MMM}
M_1=\left(\begin{array}{cc}1&1/\bar w_1\\0&1\end{array} \right)\,,\ \ \ \ \ \ \ \
\ \ M_{N+1}=\left(\begin{array}{cc}1& 1/\bar w_{N+1}\\0&1\end{array} \right)\,,
\ee
with $\bar w_{N+1}\neq \bar w_1$. They are given by linear combinations of the
operators $a(u),\ldots,d(u)$, Eq.~\re{Tc}, with the coefficients depending on the
gauge parameters $\bar w_{N+1}$ and $\bar w_1$, which are identified as the right
arguments of the kernel $Y_u(z_1,\ldots,z_N|\bar w_1,\ldots,\bar w_{N+1})$.
Similar relations hold between the entries of the monodromy matrices
$\widehat{\mathbb{T}}_N(u)$ and $\widetilde{\widehat{\mathbb{T}}}_N(u)$,
Eqs.~\re{Mo} and \re{nTo}, so that
\be\label{tADB}
\widetilde A(u;\bar w_1)+\widetilde D(u;\bar w_1)=A(u)+D(u)\,,\qquad \widetilde
B(u;\bar w_1) = B(u) + \mathcal{O}(1/\bar w_1)\,,
\ee
where we indicated explicitly the dependence on the gauge parameter $\bar w_1$.
Eqs.~\re{TYc} play a crucial r\^ole in our subsequent analysis. Their derivation
can be found in~\cite{SD,SoV}.

To proceed further let us express the  entries of the monodromy matrix of the
open chain, $\widetilde B(u)$ and $\widetilde D(u)$, in terms of those for the
closed spin chain, $\widetilde a(u),\ldots, \widetilde d(u)$. One finds from
\re{Mo}, \re{Tc} and \re{nTo}
\be
\widetilde B(u)=\widetilde b\lr{-u}\widetilde a\lr{u} -\widetilde
a\lr{-u}\widetilde b\lr{u}\,,\qquad \widetilde D(u)=\widetilde d\lr{-u}\widetilde
a\lr{u} -\widetilde c\lr{-u}\widetilde b\lr{u}\,.
\label{B}
\ee

\subsection{Kernel of the $\mathbb{Q}-$operator}

We now turn to constructing the kernel of the Baxter $\mathbb{Q}-$operator and
consider the following auxiliary operator $\mathbb{G}(u,v): \mathcal{V}_N
\longmapsto \mathcal{V}_N$ with the kernel given by the convolution of two
$Y-$functions introduced in the subsection~\ref{GT}
\ba
&&G_{u,v}(z_1,\ldots,z_N|\bar w_1,\ldots,\bar w_N)= \e^{i\pi s(2N-1)}\,\int
\mathcal{D} y_2\ldots \int \mathcal{D} y_{N}\,
\label{G-def}
\\
&&\hspace*{30mm} \times\ Y_{u}(z_1,\ldots,z_N|\bar w_1,\bar y_2,\ldots,\bar y_N,
\bar w_N)\, Y_{v}(y_2,\ldots,y_N|\bar w_1,\bar w_2,\ldots, \bar w_N)\,.
\nonumber
\ea
Here the integration measure $\mathcal{D} y_n$ is defined in \re{measure} and the
prefactor is introduced for the later convenience. Notice that two $Y-$functions
in \re{G-def} have a different number of arguments and depend on the same
variables $\bar w_1$ and $\bar w_{N}$.

\begin{figure}[t]
\psfrag{z1}[cc][cc]{$z_1$}\psfrag{z2}[cc][cc]{$z_2$}\psfrag{zn}[cc][cc]{$z_N$}
\psfrag{w1}[cc][cc]{$\bar w_1$}\psfrag{w2}[cc][cc]{$\bar w_2$}
\psfrag{wn}[cc][cc]{$\bar w_N$} \psfrag{dots}[cc][cc]{$\Mybf{\cdots}$}
\psfrag{a}[cc][cc]{$\beta_u$} \psfrag{b}[cc][cc]{$\alpha_u$}
\psfrag{c}[cc][rc]{$\beta_v$} \psfrag{d}[cc][rc]{$\alpha_v$}
\centerline{\epsfxsize10.0cm\epsfbox{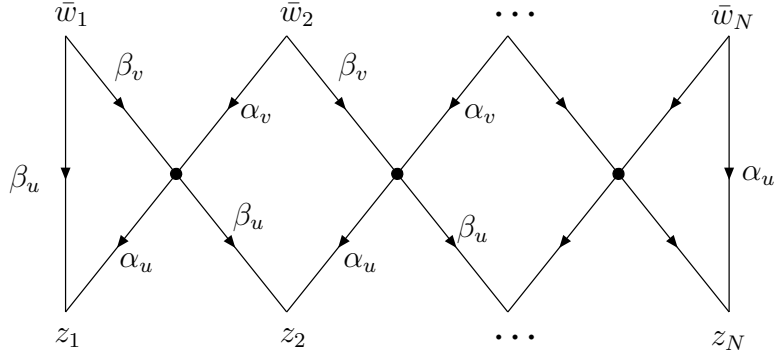}} \vspace*{0.5cm}
\caption[]{Diagrammatical representation of the function $G_{u,v}(\vec z | \vec
w)$. For $v=-u$ the diagram defines the kernel of the Baxter
$\mathbb{Q}-$operator, Eq.~\re{Q-explicit}.
Here $\alpha_x=s-ix$ and $\beta_x=s+ix$.}%
\label{fig1}%
\end{figure}

Let us demonstrate that for $v=-u$ the operator $\mathbb{G}(u,v)$ satisfies the
relations \re{Q-comm} -- \re{Bax-eq} and, therefore, it can be identified as the
Baxter $\mathbb{Q}-$operator for the homogeneous open Heisenberg spin chain
\be
\mathbb{Q}(u)= \mathbb{G}(u,-u)
\,,
\label{Q-kern}
\ee
or equivalently
\ba
&&\hspace*{-20mm} Q_u(z_1,\ldots,z_N|\bar w_1,\ldots,\bar w_N)
= \e^{i\pi s(2N-1)}\, (z_1-\bar w_1)^{-\beta_u} (z_N-\bar w_N)^{-\alpha_u}
\nonumber
\\
&& \times 
\prod_{n=2}^{N} \int \mathcal{D} y_n\, (z_{n-1}-\bar y_n)^{-\alpha_u}(z_n-\bar
y_n)^{-\beta_u} (y_n-\bar w_{n-1})^{-\alpha_u}(y_n-\bar w_n)^{-\beta_u}
\label{Q-explicit}
\ea
with $\alpha_u=s-iu$ and  $\beta_u=s+iu$. To prove \re{Q-kern} we apply the
diagrammatical approach developed in Ref.~\cite{SoV}. In this approach, one
represents the kernel $G_{u,v}(\vec{z}|\vec{w})$, Eq.~\re{G-def}, as the Feynman
diagram shown in Fig.~\ref{fig1}. There, the arrow with the index $\alpha$ that
goes from $y$ to $z$ represents the factor $(z-\bar y)^{-\alpha}$ (see
Eq.~\re{alpha-rep}) while the black blob denotes integration over the position
$w$ of the corresponding vertex with the $SL(2,\mathbb{R})$ measure
$\mathcal{D}w$ (see Eq.~\re{chain-h} and Fig.~\ref{Chain}).

The operator $\mathbb{G}(u,v)$ is symmetric under interchange of the spectral
parameters
\be\label{Guv}
\mathbb{G}(u,v)=\mathbb{G}(v,u) \,,
\ee
or equivalently $G_{u,v}(\vec{z}|\vec{w})~=~G_{v,u}(\vec{z}|\vec{w})$. The proof
of \re{Guv} is based on the permutation identity shown in Fig.~\ref{comm-f}.
Writing $\beta_u=\beta_v + i(u-v)$, one replaces the left-most vertical line in
the left diagram in Fig.~\ref{fig1} by two lines with the indices $\beta_v $ and
$i(u-v)$. Then, one displaces the line with the index $i(u-v)$ across the diagram
to the right with a help of the permutation identity until it merges with the
right-most vertical line and changes its index to $\alpha_u+i(u-v)=\alpha_v$ (see
Ref.~\cite{SoV} for details). The resulting diagram coincides with the original
one but with the spectral parameters interchanged. Furthermore, it follows from
\re{G-def} and \re{TYc} that
\ba
\widetilde a(u; \bar w_1,\bar w_{N})\,{G}_{u,v}(\vec z; \vec w) &=& (u+is)^N
{G}_{u+i,v}(\vec z; \vec w)\,,\nonumber
\\
\widetilde d(u; \bar w_1,\bar w_{N})\,{G}_{u,v}(\vec z; \vec w) &=&
(u-is)^N{G}_{u-i,v}(\vec z; \vec w)\,,
\label{G-prop}
\\
\widetilde b(u; \bar w_1,\bar w_{N})\,{G}_{u,v}(\vec z; \vec w) &=& 0\,,
\nonumber
\ea
where $\vec z=(z_1,\ldots,z_N)$ and $\vec w=(\bar w_1,\ldots,\bar w_N)$. Here the
operators $\widetilde a(u),\ldots,\widetilde d(u)$ depend on the gauge parameters
$\bar w_1$ and $\bar w_N$, which coincide with the corresponding arguments of the
kernel $G_{u,v}(z_1,\ldots,z_N|\bar w_1,\ldots,\bar w_N)$.
%

Let us demonstrate that the operator $\mathbb{Q}(u)$ satisfies the Baxter
equation \re{Bax-eq}. To this end, one examines the expression entering the
l.h.s.\ of the Baxter equation \re{Bax-eq} and applies Eqs.~\re{th}, \re{tADB}
and \re{Q-kern} to get
\be
\widehat t_N(u)\,\mathbb{Q}(u)= (\widetilde A(u) +\widetilde
D(u))\,\mathbb{G}(u,-u) = \left[\frac{2u-i}{2u}\,\widetilde
D(u)+\frac{2u+i}{2u}\,\widetilde D(-u)\right]\mathbb{G}(u,-u)\,.
\label{tQ=G}
\ee
Taking into account Eqs.~\re{B}, \re{G-prop} and \re{Guv}, one finds
\ba
\widetilde D(u)\,\mathbb{G}(u,-u) &=& \widetilde d\lr{-u} \widetilde a\lr{u}
\mathbb{G}(u,-u)
=(u+is)^N\widetilde d\lr{-u}\mathbb{G}(-u,u+i)
\nonumber\\
&=&(u+is)^N(-u-is)^N\mathbb{G}(-u-i,u+i)
\label{DG}
\ea
Substituting \re{DG} into \re{tQ=G} one concludes that the operator
$\mathbb{Q}(u)$ defined in \re{Q-explicit} verifies the Baxter relation
\re{Bax-eq}. In addition, one deduces from \re{B} and \re{G-prop} that the kernel
of the $\mathbb{Q}-$operator is nullified by the operator $\widetilde B(u)$
\be\label{Bnul}
\widetilde B(u;w_1)\, Q_u(z_1,\ldots,z_N|\bar w_1,\ldots,\bar w_N)~=~0\,.
\ee
We will use this property in Sect.~\ref{SoV} to construct the unitary
transformation to the SoV representation.

The operator $\mathbb{Q}(u)$
has the following properties
\begin{itemize}
\item Parity:
\be\label{parity}
\mathbb{Q}(u)=\mathbb{Q}(-u)\,.
\ee
\item Normalization:
\be\label{KK}
\mathbb{Q}(\pm is)=\mathbb{K}\,.
\ee
\item Hermiticity:
\be\label{hermiticity}
      \lr{\mathbb{Q}(u)}^\dagger=\mathbb{Q}(u^*)\,.
\ee
\item $SL(2)$ invariance:
\be\label{SL2-inv}
 [\mathbb{Q}(u), \vec S] =0\,,
\ee
\end{itemize}
where $\mathbb{K}$ is the unit operator on the Hilbert space of the model
$\mathcal{V}_N$~(see Eq.~\ref{K-ker} in Appendix~\ref{Ap}) and $\vec
S=\sum_{k=1}^N \vec S_k$ is the operator of the total spin. Eq.~\re{parity} is a
consequence of \re{Q-kern} and \re{Guv}. Eq.~\re{KK} follows from the fact that
$\beta_{is}=\alpha_{-is}=0$ so that the corresponding lines in the diagram in
Fig.~\ref{fig1} disappear leading to drastic simplification of the kernel.
Eq.~\re{hermiticity} follows directly from the definition of the conjugated
operator $\lr{\mathbb{Q}(u)}^\dagger$. To verify \re{SL2-inv} one notices that
the kernel of the $\mathbb{Q}-$operator, Eq.~\re{Q-explicit}, is transformed
under the $SL(2,\mathbb{R})$ transformations as
\be
{Q}_u(\vec z'|\vec w') = \prod_{k=1}^N (\gamma \bar w_k + \delta)^{2s} (\gamma
z_k + \delta)^{2s}\,{Q}_u(\vec z|\vec w)
\label{Q-trans}
\ee
where $z_k'=(\alpha z_k+\beta)/(\gamma z_k+\delta)$ and $\bar w_k'=(\alpha\bar
w_k+\beta)/(\gamma\bar w_k+\delta)$ with real $\alpha,\ldots,\delta$ such that
$\alpha\delta-\beta\gamma=1$. 

\begin{figure}[t]
\psfrag{z1}[cc][cc]{$z_1$}\psfrag{z2}[cc][cc]{$z_2$}\psfrag{zN}[cc][cc]{$z_N$}
\psfrag{w1}[cc][cc]{$\bar w_1$}\psfrag{w2}[cc][cc]{$\bar w_2$}
\psfrag{wN}[cc][cc]{$\bar w_N$} \psfrag{dots}[cc][cc]{$\Mybf{\cdots}$}
\psfrag{=}[cc][cc]{$\Mybf{=}$} \psfrag{a1}[cc][rc]{$\beta_v$}
\psfrag{b1}[cc][rc]{$\alpha_v$} \psfrag{a2}[cc][rc]{$\beta_u$}
\psfrag{b2}[cc][rc]{$\alpha_u$}

\centerline{\epsfxsize17.0cm\epsfbox{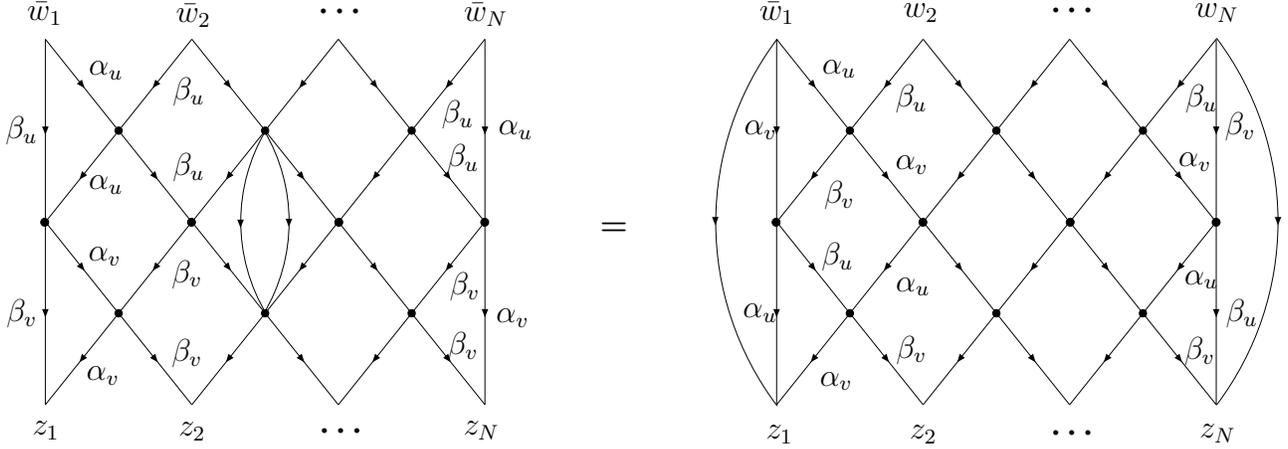}} \vspace*{0.5cm}
\caption[]{Diagrammatical proof of Eq.~\re{Identity}. The left diagram represents
the kernel of the operator $\mathbb{Q}(v)\mathbb{Q}(u)$. The right diagram is
obtained by displacing two wavy lines carrying the indices $\pm i(u+v)$ to the
right/left with a help of the permutation identity. Here $\alpha_x=s-ix$,
$\beta_x=\alpha_{-x}=s+ix$. }
\label{fig2}
\end{figure}

We are now ready to demonstrate that the $\mathbb{Q}-$operator \re{Q-kern}
satisfies the relations \re{tQ} and \re{Q-comm}. To verify \re{tQ}, one performs
the Hermitian conjugation of the both sides of the Baxter equation \re{Bax-eq}.
Taking into account \re{hermiticity}, one finds that the r.h.s.\ of \re{Bax-eq}
goes into $\widehat t_N(u^*)\mathbb{Q}(u^*)$ whereas its l.h.s.\ is replaced by
$\lr{ \widehat t_N(u)\mathbb{Q}(u)}^\dagger=\mathbb{Q}(u^*) \widehat t_N(u^*)$.
Equating the two expressions one arrives at \re{tQ}. Finally, let us show that
the operator \re{Q-kern} satisfies the commutativity condition \re{Q-comm}. The
proof can be performed diagrammatically. To this end, one examines the Feynman
diagram corresponding to the product
$\mathbb{Q}(v)\mathbb{Q}(u)=\mathbb{G}(v,-v)\,\mathbb{G}(u,-u)$ and inserts a
pair of lines with the indices $\pm i(u+v)$ into one of the central rhombuses as
shown in Fig.~\ref{fig2}. Displacing the two lines horizontally in the opposite
directions with a help of the permutation identity (see Fig.~\ref{comm-f}) one
obtains the Feynman diagrams shown in Fig.~\ref{fig2} to the right. It differs
from the original diagram in that various $\alpha-$ and $\beta-$indices got
interchanged and two additional lines with the indices $\pm i(u+v)$ connect the
``end points'', $\bar w_1$ with $z_1$ and $\bar w_N$ with $z_N$. Taking into
account the definition of the $\mathbb{G}-$operator, Eq.~\re{G-def} (see
Fig.~\ref{fig1}) one finds that the Feynman integral corresponding to this
diagram can be written as
\be
\bigg[\mathbb{Q}(v)\,\mathbb{Q}(u)\bigg](\vec z; \vec w) ={(z_1-\bar
w_1)^{i(u+v)} (z_N-\bar w_N)^{-i(u+v)}} \left[\lr{\mathbb{G}(u^*,v^*)}^\dagger
\,\mathbb{G}(-u,-v)\right](\vec z; \vec w)\,,
\label{Identity}
\ee
where the kernel of the integral operator $\lr{\mathbb{G}(u^*,v^*)}^\dagger$ is
given by $(G_{u^*,v^*}(w_1,\ldots,w_N|\bar z_1,\ldots,\bar z_N))^*$. According to
\re{Guv}, the r.h.s.\ of \re{Identity} is invariant under interchanging
$u\leftrightarrows v$ thus proving the commutativity relation \re{Q-comm}.

\subsection{Contour-integral representation for the $\mathbb{Q}-$operator}

In the previous subsection we constructed the Baxter $\mathbb{Q}-$operator for
the homogeneous open spin chain, Eqs.~\re{Q-explicit}. As was already mentioned,
the $\mathbb{Q}-$operator is diagonalized by the eigenstates of the model,
Eq.~\re{Bax-eig}, and the corresponding eigenvalues $Q_{\mybf{q}}(u)$ satisfy
\re{Bax-eq}.

The Baxter equation \re{Bax-eq} is a finite-difference functional equation and
its solutions are defined up to multiplication by an arbitrary periodic function,
$f(u+i)=f(u)$. To fix this ambiguity and determine eigenvalues of the
$\mathbb{Q}-$operator, one has to specify analytical properties of
$Q_{\mybf{q}}(u)$.
They can be identified using the following contour-integral representation for
the $\mathbb{Q}-$operator on the quantum space of the model
$\Psi(z_1,\ldots,z_N)\in \mathcal{V}_N$
\ba\label{Q-cont}
&&\left[\mathbb{Q}(u)\Psi\right](z_1,\ldots,z_N)=[B(s+iu,s-iu)]^{-2N+1}\\[3mm]
&&\ \ \ \times \int_0^1\prod_{n=1}^N d\sigma_n\,
(1-\sigma_n)^{s+iu-1}\,\sigma_n^{s-iu-1} \int_0^1\prod_{k=2}^N d\tau_k\,
(1-\tau_k)^{s+iu-1}\,\tau_k^{s-iu-1}\, \Psi(Z_1,\ldots,Z_N)\,,\nonumber
\ea
where $B(x,y)$ is the Euler beta-function and the $Z-$coordinates are defined as
\ba\label{Z}
&Z_1&=(1-\sigma_1) z_1+\sigma_1[\tau_2 z_1+(1-\tau_2) z_2]\,\nonumber\\
&Z_k&=(1-\sigma_k)[\tau_k z_{k-1}+(1-\tau_k) z_k]~+~
\sigma_k [\tau_{k+1} z_{k}+(1-\tau_{k+1}) z_{k+1}]\,, \ \  \  \, (1<k<N)\,,\nonumber\\
&Z_N&=(1-\sigma_N)[\tau_N z_{N-1}+(1-\tau_N) z_N]~+~\sigma_N z_{N}\,.
\ea
To obtain \re{Q-cont}, one uses the integral representation for the
$\mathbb{Q}-$operator, Eq.~\re{Q-explicit}, and applies the identity \re{drift}.

Since the function $\Psi(Z_1,\ldots,Z_N)$ is holomorphic in the upper half-plane
$\Im z_n>0$, the integral in the r.h.s.\ of \re{Q-cont} is convergent inside the
strip $-s < \Re (iu) < s$ in the complex $u-$plane. Analytically continuing the
integral outside this strip, one finds that it contains poles of the order $p \le
2N-1$ originating from integration at the vicinity of the end-points $\sigma_n
\,, \tau_k\to 0,1$. They are located at $iu = \pm (n-s)$ (with $n$ nonnegative
integer) and are compensated by the beta-function prefactor entering \re{Q-cont}.
As a result, $[\mathbb{Q}(u)\Psi](z_1,\ldots,z_N)$ does not have poles in $u$
and, therefore, the eigenvalues of the Baxter operator, $Q_{\mybf{q}}(u)$, are
entire functions of $u$.

Eq.~\re{Q-cont} also allows one to determine asymptotic behaviour of
$Q_{\mybf{q}}(u)$ at large $u$. It is given by
\be
Q_{\mybf{q}}(u) \sim u^{2h} \left[1+ \mathcal{O}(1/u^2)\right],
\label{Q-asym}
\ee
with $h$ nonnegative integer defining the total spin of the model
\be
[\vec S^2-(h+Ns)(h+Ns-1)]\Psi_{\mybf{q}}(\vec z)=0\,.
\label{S2-eig}
\ee
To establish \re{Q-asym}, one verifies using \re{Q-cont} that the Baxter operator
is invariant under arbitrary $SL(2,\mathbb{C})$ transformations, in particular
under the following one $z\mapsto -i\lr{w+i}/\lr{w-i}$
\be\label{RtU}
\Psi(z_1,\ldots,z_N)\mapsto \widetilde\Psi(w_1,\ldots,w_N)=\prod_{k=1}^N
(w_k-i)^{-2s}\Psi\left(-i\frac{w_1+i}{w_1-i},\ldots,-i\frac{w_N+i}{w_N-i}\right)\,,
\ee
which map the upper half-plane $\Im z_k >0$ into a unit disk $|w_k|<1$. The main
advantage of dealing with functions $\widetilde\Psi(w_1,\ldots,w_N)$ holomorphic
inside the unit circle is that solutions to \re{S2-eig} have a simple form in
that case. Namely, the Hilbert space of the model contains the highest weigths
which satisfy \re{S2-eig} and are given by homogeneous translation invariant
polynomials of degree $h$,
$\widetilde\Psi(w_1,\ldots,w_N)=P_h(w_1-w_2,\ldots,w_{N-1}-w_N)$. Since the
Baxter operator \re{Q-cont} remains invariant under \re{RtU}, one can substitute
the function $\Psi(\vec z)$ in \re{Q-cont} by such polynomial. Then,
$\Psi(Z_1,\ldots,Z_N)$ entering \re{Q-cont} becomes a polynomial in the $\sigma-$
and $\tau-$parameters. Integrating term-by-term in the r.h.s.\ \re{Q-cont} one
finds that the dominant contribution at large $u$ comes from terms containing a
maximum number of $\sigma$'s and $\tau$'s. This number equals $2h$ and leads to
the asymptotics \re{Q-asym}.

Given that $Q_{\mybf{q}}(u)$ is an even function of $u$, Eq.~\re{parity}, and
making use of \re{hermiticity}, we conclude that the eigenvalues of the
$\mathbb{Q}-$operator are real polynomials in $u^2$ of degree $h$
\be
Q_{\mybf{q}}(u) = a_{\mybf{q}}\prod_{k=1}^h (u^2-\lambda_k^2)\,, \qquad
\lr{Q_{\mybf{q}}(u)}^*= Q_{\mybf{q}}(u^*)
\label{roots}
\ee
with the normalization constant $a_{\mybf{q}}$ fixed by the condition
$Q_{\mybf{q}}(is)=1$, Eq.~\re{KK}. Substituting \re{roots} into \re{Bax-eq} and
putting $u=\lambda_k^2$, one finds that the roots $\lambda_k^2$ satisfy the Bethe
equations for the open spin chain~\cite{Sklyanin88}.

\subsection{Relation to the Hamiltonian}

Let us demonstrate that the Hamiltonian of the open spin chain, Eq.~\re{H}, is
given by a logarithmic derivative of the Baxter operator evaluated at special
values of the spectral parameter $u=\pm is$. Due to \re{KK} the expansion of the
$\mathbb{Q}-$operator around $u=\pm is$ can be written as
\be\label{Q-H}
\left[\mathbb{Q}(\pm is
  +\epsilon)\Psi\right](\vec{z})~=~\Psi(\vec{z})~\mp~
i\epsilon\, \left[\CH_N\,\Psi\right](\vec{z})~+~{\cal O}(\epsilon^2)\,,
\ee
with $\CH_N$ being some integral operator. Its explicit form can be found from
the contour-integral representation for the $\mathbb{Q}-$operator, \re{Q-cont}.
At $u=-is+\epsilon$ the beta-prefactor in the r.h.s.\ of \re{Q-cont} vanishes as
$\epsilon^{2N-1}$ but it is compensated by poles coming from integration at the
vicinity of $\sigma_n=\tau_k =0$. Carefully separating contribution from this
region, one obtains that the operator $\CH_N$ entering \re{Q-H} is given by the
sum of two-particle integral operators
\be\label{HH}
\CH_N =-i\frac{d}{d\epsilon}\ln \mathbb{Q}(- is+\epsilon)\biggl|_{\epsilon=0}=
\sum_{n=0}^{N-1} H_{n,n+1}\,,
\ee
where $H_{n,n+1}$ acts on $\Psi(z_n,z_{n+1})\in V_n\otimes V_{n+1}$ as
\be\label{Hn}
[H_{12}\,\Psi](z_1,z_2)~=~-\int_0^1\frac{d\tau}{\tau}  {{(1-
    \tau)}^{2s-1}}
\left[\Psi(z_{12}(\tau),z_2)+\Psi(z_1,z_{21}(\tau))-2\Psi(z_1,z_2) \right]\,,
\ee
with $z_{ik}(\tau)=(1-\tau)z_i+\tau z_k$. It is straightforward to check that the
Hamiltonian $H_{n,n+1}$ commutes with the two-particle spin
$\vec{S}_n+\vec{S}_{n+1}$ defined in \re{s-rep} and, therefore, it only depends
on the Casimir operator $J_{n,n+1}$, Eq.~\re{JJ}. To find the explicit form of
this dependence one applies $H_{12}$ to the state $\Psi(z_1,z_2)=
(z_1-z_2)^h/((z_1+i)(z_2+i))^{h+2s}$ with $h$ nonnegative
integer.%
\footnote{Under conformal mapping \re{RtU} this state is transformed into a
homogeneous polynomial of degree $h$, $\widetilde\Psi(w_1,w_2)=(w_1-w_2)^h$.} It
diagonalizes simultaneously the Casimir operator $J_{12}\,\Psi = (h+2s)\Psi$ and
the two-particle kernel $H_{12} \Psi= 2[\psi(h+2s)-\psi(2s)]\Psi$ leading to the
expression
\be\label{Hnn}
H_{n,n+1}~=~2\left[\psi(J_{n,n+1})~-~\psi(2s)\right]\,,
\ee
which coincides with \re{H}.

Eq.~\re{HH} establishes the relation between the Hamiltonian of the model \re{H}
and the Baxter $\mathbb{Q}-$operator. Obviously, the same relation holds between
their eigenvalues
\be\label{QE}
E_{\mybf{q}}~=~ \pm i\frac{d}{d\epsilon}\ln Q_{\mybf{q}}(\pm
is+\epsilon)\biggl|_{\epsilon=0}\,.
\ee
Thus, to reconstruct the energy spectrum of the model, one has to find polynomial
solutions to the Baxter equation \re{Bax-eq} and apply \re{QE}.

\setcounter{equation}{0}
\section{Separation of Variables}
\label{SoV}


In this section we will construct integral representation for the eigenstates of
the model, Eq.~\re{Sch}, by going over to the representation of the Separated
Variables $(p,\Mybf{x})=(p,x_1,...,x_{N-1})$ (SoV)
\be
\Psi_{\mybf{q},p}\,(z_1,\ldots,z_N)=\int_{\mathbb{R}_{+}^{N-1}}
d^{N-1}\Mybf{x}\,\mu(\Mybf{x})\, U_{p,\mybf{x}}(z_1,\ldots,z_N)\,
\Phi_{\mybf{q}}(\Mybf{x})\,.
\label{SoV-gen}
\ee
Here $\Phi_{\mybf{q}}(\Mybf{x})$ is the eigenfunction of the model in the
separated variables. It is factorized into a product of functions depending on a
single variable $\Phi_{\mybf{q}}(\Mybf{x})\sim Q(x_1)\ldots Q(x_{N-1})$. As will
be shown in this section, $Q(x_k)$ coincides with the eigenvalue of the Baxter
$\mathbb{Q}-$operator. The kernel $U_{p,\mybf{x}}$ of the unitary operator
corresponding to the SoV transformation is defined as
\be
U_{p,\mybf{x}}(z_1,\ldots,z_N)=\vev{z_1,\ldots,z_N|p,\Mybf{x}}\,.
\label{U-ker}
\ee
We will argue below that the separated variables $x_k$ (with $k=1,\ldots,N-1$)
take real positive values so that integration in \re{SoV-gen} goes over
$\Mybf{x}\in \mathbb{R}_+^{N-1}$ with $d^{N-1}\Mybf{x}=d x_1 \ldots dx_{N-1}$ and
$\mu(\Mybf{x})$ being a nontrivial integration measure. Eq.~\re{SoV-gen} defines
the transformation $\Phi_{\mybf{q}}\mapsto\Psi_{\mybf{q},p}$. The inverse
transformation looks as follows
\be
\Phi_{\mybf{q}}(\Mybf{x})\,\delta(p-p') = \vev{p',\Mybf{x}|\Psi_{\mybf{q},p}}
=\int {\cal D}^N z\, \lr{U_{p',\mybf{x}}
(z_1,\ldots,z_N)}^*\Psi_{\mybf{q},p}\,(z_1,\ldots,z_N)\,.
\label{SoV-inv}
\ee

To construct the unitary transformation to the SoV representation one has to
specify the complete set of the states $\ket{p,\Mybf{x}}$ and define the
corresponding kernel \re{U-ker}. It is well-known that for the $SL(2)$ spin chain
with periodic boundary conditions, within the framework of the Sklyanin's
approach~\cite{Sklyanin}, the basis vectors $\ket{p,\Mybf{x}}$ can be defined as
eigenvectors of the operator $b(u)$ which is the off-diagonal matrix element of
the monodromy matrix $T_N(u)$, Eq.~\re{Tc}. We will demonstrate that the same
recipe also works for the open spin chain. Namely, the basis vectors
$\ket{p,\Mybf{x}}$ in \re{U-ker} can be defined as the eigenstates of the
operator $B(u)$ entering the expression for monodromy matrix $\mathbb{T}_N(u)$,
Eq.~\re{nTo}.

According to \re{nTo}, $B(u)$ is a polynomial in $u$ of degree $2N-1$ with
operator-valued coefficients, $B(u) = 2i(-1)^{N-1} S_- u^{2N-1}+\ldots$. In
addition, it follows from \re{A-minus} that $B(-i/2)=0$ and, moreover,
$B(u)/(2u+i)$ is an even function of $u$. This suggests to remove the
``kinematic'' zero of $B(u)$ and define the operator
\be\label{hatB}
\widehat B(u) ~=~\frac{B(u)}{2u+i}=(-1)^{N-1}iS_{-}\, \left(u^{2N-2}~+~\widehat
b_2\,u^{2N-4}~+~\ldots~+~\widehat b_N \right),
\ee
with $\widehat b_2,\ldots,\widehat b_N$ being some (commuting) operators. Since
$B(u)=b(-u)a(u)-a(-u)b(u)$ (see Eq.~\re{B}), one finds using $\lr{a(u)}^\dagger =
a(u^*)$ and $\lr{b(u)}^\dagger = b(u^*)$ that $\lr{B(u)}^\dagger=-B(-u^*)$, or
equivalently $(\widehat B(u))^\dagger= \widehat B(u^*)$. Thus, $\widehat B(u)$ is
hermitian operator for real $u$.

Following Sklyanin~\cite{Sklyanin}, we identify the eigenstates of the operator
$\widehat B(u)$ as the kernel of the transition operator to the SoV
representation
\be
\widehat{B}(u)\,U_{p,\mybf{x}}(z_1,\ldots,z_N)=(-1)^{N-1}\,p\,(u^2-x_1^2)\cdots(u^2-x_{N-1}^2)\,
U_{p,\mybf{x}}(z_1,\ldots,z_N)\,.
\label{B-pol}
\ee
According to \re{BB}, $[\widehat{B}(u),\widehat{B}(v)]=0$ and, therefore,
$U_{p,\mybf{x}}(z_1,\ldots,z_N)$ does not depend on the spectral parameter $u$.
Due to \re{hatB}, the corresponding eigenvalues are real polynomials in $u^2$ of
degree $N-1$. They can be parameterized by the total momentum $p$ and by the set
of parameters $\Mybf{x}=(x_1,\ldots,x_{N-1})$ which are identified as the
separated variables. Hermiticity of the operator $\widehat{B}(u)$ implies that
$x_k^2$ can be either real, or can appear in complex conjugated pairs,
$x_k^2=(x_j^2)^*$. We will argue in the next section that the separated variables
satisfy a much stronger condition $x_k^2>0$, which together with the symmetry of
\re{B-pol} under $x_k\to -x_k$ allows one to assign to the separated variables
$\Mybf{x}$ \textit{real positive} values. This follows from the requirement that
$U_{p,\mybf{x}}(z_1,\ldots,z_N)$ have to be the eigenstates of the self-adjoint
operator $\widehat{B}(u)$ and, therefore, they have to fulfill the completeness
condition
\be
\int_0^\infty dp \int_{\mathbb{R}_+^{N-1}}
d^{N-1}\Mybf{x}\,\mu(\Mybf{x})\,\lr{U_{p,\mybf{x}}(w_1,\ldots,w_N)}^*\,U_{p,\mybf{x}}(z_1,\ldots,z_N)
=\prod_{n=1}^N \mathbb{K}(z_n|\bar w_n)
\label{complete}
\ee
where $\mathbb{K}(z |\bar w)=\e^{i\pi s} (z-\bar w)^{-2s}$ is the kernel of the
identity operator (see Eq.~\re{K-ker}).

The diagonal element $D(\pm x_k)$ of the monodromy matrix \re{nTo} acts on
$U_{p,\mybf{x}}(w_1,\ldots,w_N)$ as a shift operator
\be\label{D-shift}
D(\pm x_k)U_{p,\mybf{x}}(z_1,\ldots,z_N) = \delta(\pm x_k) U_{p,\mybf{x}\pm
i\mybf{e}_k}(z_1,\ldots,z_N)\,.
\ee
Indeed, taking $v=\pm x_k$ in the second fundamental relation \re{BB} and
applying its both sides to $U_{p,\mybf{x}}(z_1,\ldots,z_N)$, one arrives at
\re{D-shift}. The scalar factor $\delta(x_k)$ depends on the normalization of
$U_{p,\mybf{x}}(z_1,\ldots,z_N)$. Applying $U_{p,\mybf{x}}$ to the both sides of
\re{q-det} and taking $u=x_k$ one finds that $\delta(x_k)$ satisfies the relation
\be
\delta(x_k)\delta(-x_k-i)=[(x_k+is)(x_k+i(1-s))]^{2N}\,.
\ee
In \re{D-shift} it was tacitly assumed that the function
$U_{p,\mybf{x}}(z_1,\ldots,z_N)$ can be continued to complex $\Mybf{x}$. Notice
that $U_{p,\mybf{x}\pm i\mybf{e}_k}(z_1,\ldots,z_N)$ is not the eigenfunction of
the operator $\widehat B(u)$ even though it satisfies the differential equation
\re{B-pol}.

\subsection{Transition kernel}

\begin{figure}[t]
\psfrag{z1}[cc][cc]{$z_1$}\psfrag{z2}[cc][cc]{$z_2$}\psfrag{zn}[cc][cc]{$z_N$}
\psfrag{w2}[cc][cc]{$\bar w_2$} \psfrag{wn}[cc][cc]{$\bar w_N$}
\psfrag{dots}[cc][cc]{$\Mybf{\cdots}$} \psfrag{a}[cc][cc]{$\beta_x$}
\psfrag{b}[cc][cc]{$\alpha_x$} \psfrag{c}[cc][rc]{$\alpha_x$}
\psfrag{d}[cc][rc]{$\beta_x$} \centerline{\epsfxsize10.0cm\epsfbox{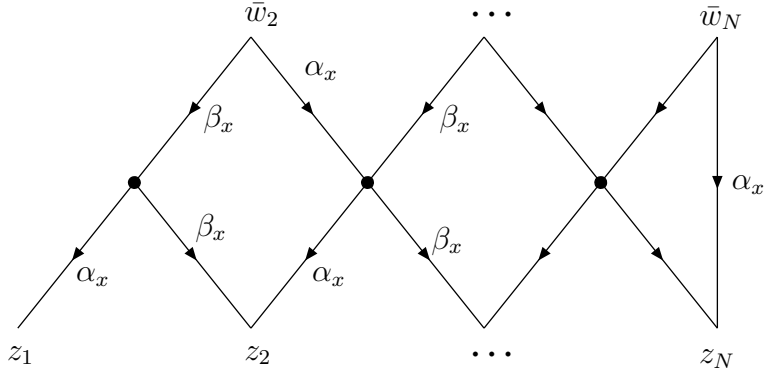}}
\vspace*{0.5cm} \caption[]{ The diagrammatical representation of the function
$\Lambda_{x}(z_1,\ldots,z_N|\bar w_2,\ldots,\bar w_N)$.
}%
\label{lamo}%
\end{figure}%

Solving the spectral problem \re{B-pol}, we follow the approach developed in
Ref.~\cite{SoV} in application to the closed spin chain. To begin with, we notice
that the differential equation \re{B-pol} is equivalent to the system of $N$
equations
\begin{equation}
iS_-\,U_{p,\mybf{x}}(z) = p\,U_{p,\mybf{x}}(z)\,,\qquad \widehat{B}(\pm
x_k)\,U_{p,\mybf{x}}(z)=0,\qquad (k=1,...N-1)\,.
\label{BxN}
\end{equation}
Let us consider the second relation and compare it with a similar relation
\re{Bnul} for $u=\pm x_k$. Sending the gauge parameter $w_1$ in \re{Bnul} to
infinity and taking into account \re{tADB} one finds that $B(\pm x_k)$
annihilates the following function 
\be\label{Lambda}
\Lambda_{x}(z_1,\ldots,z_N|\bar w_2\ldots,\bar w_{N})~=~ \lim_{\bar w_1\to\infty}
\bar w_1^{2s}\, Q_x(z_1,\ldots,z_N|\bar w_1,\ldots,\bar w_N)\,.
\ee
Here the additional prefactor is inserted to make the limit finite
\ba
\Lambda_{u}(z_1,\ldots,z_N|\bar w_2\ldots,\bar w_{N}) &=&\e^{i\pi s(2N-1)} \int
\mathcal{D} y_2 \ldots \mathcal{D} y_N\,(y_N-\bar w_N)^{-\beta_u}(z_N-\bar
w_N)^{-\alpha_u}
\label{Lambda-ker}
\\
&&\hspace*{-20mm}\times \prod_{k=2}^N (z_{k-1}-\bar y_k)^{-\alpha_u} (z_k-\bar
y_k)^{-\beta_u}\prod_{n=2}^{N-1} (y_n-\bar w_n)^{-\beta_u}(y_{n+1}-\bar
w_n)^{-\alpha_u}\,. \nonumber
\ea
As before, it is convenient to represent this expression as the Feynman diagram
shown in Fig.~\ref{lamo}. It differs from the Feynman diagram for the
$\mathbb{Q}-$operator (see Fig.~\ref{fig1}) in that two lines attached to the
vertex $\bar w_1$ are removed.

By the construction, the $\Lambda-$function satisfies the relation
\be\label{Bn}
\widehat{B}(\pm x_k)\,\Lambda_{x_k}(z_1,\ldots,z_N|\bar w_2\ldots,\bar
w_{N})~=~0\,.
\ee
Let us introduce the integral operator $\Lambda_N(x)$ with the kernel given by
\re{Lambda-ker}. It maps a function of $N-1$ variables
$\Psi_{N-1}(w_2,\ldots,w_N)$ into a function of $N$ variables
$\Psi_N(z_1,\ldots,z_N)$
\ba
\label{Lambda1}
\Psi_N(z_1,\ldots,z_N) &=& [\Lambda_N(u) \, \Psi_{N-1}](z_1,\ldots,z_N) \\
&=& \int \mathcal{D} w_2 \ldots \int \mathcal{D} w_N
\,\Lambda_u(z_1,\ldots,z_N|\bar w_2\ldots,\bar
w_{N})\Psi_{N-1}(w_2,\ldots,w_N)\,, \nonumber
\ea
The operator $\Lambda_N(x)$ defined in this way has a number of remarkable
properties:
\begin{itemize}
\item{Parity:}
\be\label{PL}
\Lambda_N(x)=\Lambda_N(-x)
\ee
\item{Commutativity:}
\be\label{CL}
\Lambda_N(x_1)\Lambda_{N-1}(x_2) = \Lambda_N(x_2)\Lambda_{N-1}(x_1)
\ee
\item{Baxter relation:}
\be
\widehat t_N(x)\,\Lambda_N(x)=\Delta_{+}(x)\,\Lambda_N(x+i)~+~
\Delta_{-}(x)\,\Lambda_N(x-i)
\label{Bax-Lambda}
\ee
\item{Exchange relation:}
\be
\Lambda_N^\dagger(x) \Lambda_N(y) =\varphi(x,y)\cdot
\Lambda_{N-1}(y)\Lambda_{N-1}^\dagger(x)\,,
\label{exchange}
\ee
\end{itemize}
where $x\neq y$ and the scalar function $\varphi(x,y)=\varphi(y,x)$ is defined as
\be
\varphi(x,y)=\e^{4i\pi s}\,a(\alpha_x,\alpha_y)\,a(\beta_x,\beta_y)\,
a(\beta_x,\alpha_y)\,a(\alpha_x,\beta_y)
\label{varphi}
\ee
with $\alpha_x=s-ix$ and $\beta_x=s+ix$,
\be
a(\alpha,\beta)=\e^{-i\pi s}\,
\frac{\Gamma(\alpha+\beta-2s)\Gamma(2s)}{\Gamma(\alpha)\Gamma(\beta)}\,.
\label{a}
\ee
The following comments are in order.

\begin{figure}[t]
\psfrag{=}[cc][cc]{$=$} \psfrag{a1}[cc][cc]{$\beta_x$}
\psfrag{b1}[cc][cc]{$\alpha_x$} \psfrag{a2}[cc][cc]{$\alpha_y$}
\psfrag{b2}[cc][cc]{$\beta_y$}
\centerline{\epsfxsize16.0cm\epsfbox{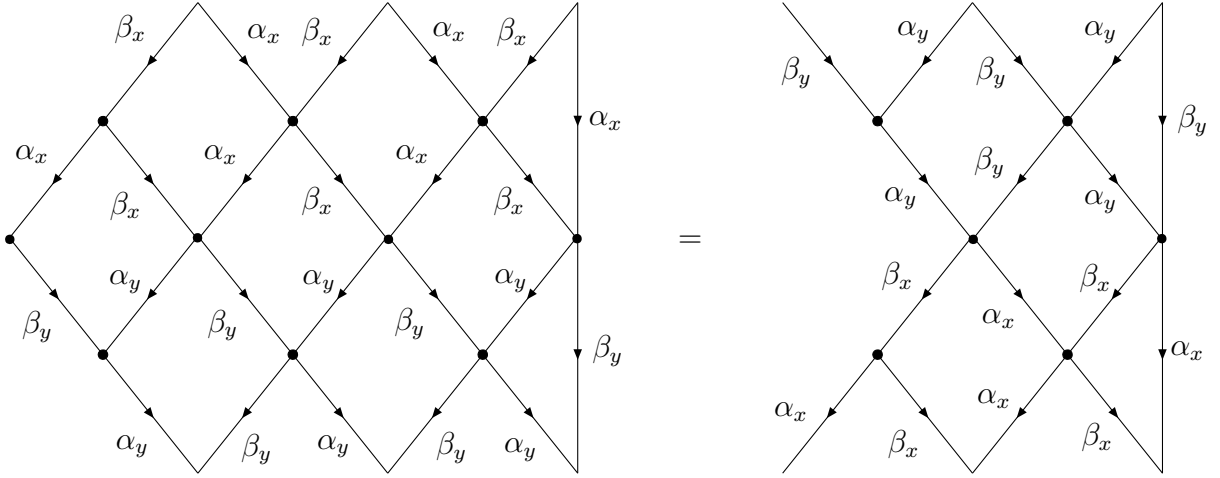}} \vspace*{0.5cm}
\caption[]{Exchange relation. Here $\alpha_x=s-ix$ and $\beta_x=s+ix$. }
\label{fig3}
\end{figure}

Eq.~\re{PL} follows from the parity property of the Baxter $\mathbb{Q}-$operator,
Eq.~\re{parity}. The proof of \re{CL} can be performed diagrammatically with a
help of the permutation identities (see Figs.~\ref{comm-f} and \ref{amp1}). It
goes along the same lines as the proof of commutativity property for the
$\mathbb{Q}-$operator presented at the end of Sect.~3.2. Eq.~\re{Bax-Lambda}
follows immediately from \re{Lambda} and \re{Bax-eq}. The proof of the exchange
relation \re{exchange} is illustrated in Fig.~\ref{fig3}. The product
$\Lambda_N^\dagger(x)\Lambda_N(y)$ corresponds to the left diagram in
Fig.~\ref{fig3}. The left-most vertex in this diagram can be integrated out with
a help of the chain relation (see Fig.~\ref{Chain}) producing a single line with
the index $\alpha_x+\beta_y-2s=-i(x-y)$. Then, one moves this line horizontally
to the right of the diagram by applying the permutation identity (see
Fig.~\ref{comm-f}). Repeating the same steps for the resulting diagram one
finally arrives at the right diagram in Fig.~\ref{fig3} with the additional
prefactor \re{varphi}.

Taking into account the properties of the $\Lambda-$operator, it becomes
straightforward to write a general solution to the system \re{BxN}
\be
U_{p,\mybf{x}}(z_1,\ldots,z_N)= p^{Ns-1/2} \int \mathcal{D}w_N\, \e^{ip\,w_N}\,
U_{\mybf{x}}(\vec z\,;\bar w_N)\,,
\label{B-ei}
\ee
where $U_{\mybf{x}}(\vec z\,;\bar w_N)$ is factorized into the product of $N-1$
operators
\be
U_{\mybf{x}}(\vec z\,;\bar w_N)=
\left[\Lambda_N(x_1)\,\Lambda_{N-1}(x_2)\ldots\Lambda_2(x_{N-1})
\right](z_1,\ldots,z_N|\bar w_N)\,,
\label{U}
\ee
with $\vec z=(z_1,\ldots,z_N)$ and the additional factor $p^{Ns-1/2}$ introduced
in \re{B-ei} for the later convenience. Indeed, the first relation in \re{BxN} is
satisfied due to invariance of \re{U} under translations $z_k \to z_k+\epsilon$
and $\bar w_N \to\bar w_N+\epsilon$ with $\epsilon$ real. It follows from \re{PL}
and \re{CL} that $U_{\mybf{x}}(\vec z\,;\bar w_N)$ is an even symmetric function
of $x_1,\ldots,x_{N-1}$. Since $\widehat{B}(\pm x_1)\,U_{\mybf{x}}(\vec z\,;\bar
w_N)=0$ by virtue of \re{U} and \re{Bn}, the second relation in \re{BxN} is
fulfilled for arbitrary $k$. Notice that the kernel $U_{\mybf{x}}(\vec z\,;\bar
w_N)$ satisfies a multi-dimensional Baxter relation
\be
\widehat t_N(x_k)\,U_{\mybf{x}}(\vec z\,;\bar w_N)=\Delta_+(x_k)
U_{\mybf{x}+i\mybf{e}_k}(\vec z\,;\bar w_N)+\Delta_-(x_k)
U_{\mybf{x}-i\mybf{e}_k}(\vec z\,;\bar w_N)\,,
\label{Bax-U}
\ee
where $\Mybf{e}_k$ denotes a unit basis vector in the $\Mybf{x}-$space,
$\Mybf{x}=\sum_k x_k \Mybf{e}_k$. Eq.~\re{Bax-U} follows from the similar
property of the $\Lambda-$operator, Eq.~\re{Bax-Lambda}, and the symmetry of the
kernel under permutations of $x-$variables.

\begin{figure}[t]
\psfrag{=}[cc][cc]{$=$} \psfrag{a1}[cc][cc]{$\beta_{x_1}$}
\psfrag{b1}[cc][cc]{$\alpha_{x_1}$} \psfrag{a2}[cc][cc]{$\beta_{x_3}$}
\psfrag{b2}[cc][cc]{$\alpha_{x_3}$} \psfrag{a3}[cc][cc]{$\beta_{x_2}$}
\psfrag{b3}[cc][cc]{$\alpha_{x_2}$}


\psfrag{z1}[cc][cc]{$z_1$}
\psfrag{z2}[cc][cc]{$z_2$}

\psfrag{z3}[cc][cc]{$z_3$} \psfrag{z4}[cc][cc]{$z_4$} \psfrag{w1}[cc][cc]{$\bar
w_4$}

\centerline{\epsfysize10.0cm\epsfbox{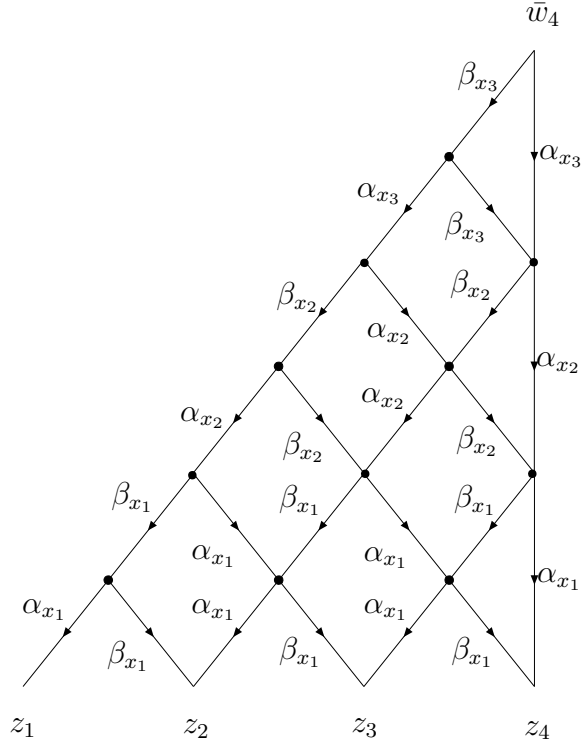}} 
\caption[]{The diagrammatic representation of the kernel
$U_{\Mybf{x}}(\vec{z},w_N)$ for $N=4$.}
\label{opyr}
\end{figure}

Eqs.~\re{B-ei}  and \re{U} define the transition kernel
$U_{p,\mybf{x}}(z_1,\ldots,z_N)$ to the SoV representation for the homogeneous
open spin chain. Remarkably enough, these expressions
have the same form as for the closed $SL(2)$ spin chain~\cite{SoV}. The only
difference between the two cases is in the definition of the $\Lambda-$operator.
Diagrammatical representation for the transition kernel \re{U} is shown in
Fig.~\ref{opyr}. The corresponding Feynman diagram has a pyramidal form which
reflects the structure of the kernel \re{U}. It consists of $(N-1)-$rows with
each row representing a single $\Lambda-$operator.

It remains to verify that the kernel \re{B-ei} satisfies for real $\Mybf{x}$ the
completeness condition \re{complete}. As we will show in Sect.~5, the
transformation to the SoV representation for $N=2$ open spin chain coincides with
the Fourier-Jacobi transform (see, e.g. Ref.~\cite{Koor}). Then, reality
condition for the separated variable $x$ and completeness condition for
$U_{p,x}(z_1,z_2)$ follow immediately from the properties of the Fourier-Jacobi
transform. For $N\ge 3$ some arguments will be presented in the next subsection.

\subsection{Integration measure}

Let us demonstrate that the transition kernel defined in Eqs.~\re{B-ei} and
\re{U} satisfies the orthogonality condition
\ba
\vev{p',\Mybf{x'}|p,\Mybf{x}} &=& \int \mathcal{D}^N z\,
U_{p,\mybf{x}}(z_1,...,z_N) \lr{U_{p',\mybf{x'}}(z_1,...,z_N)}^* \nonumber
\\
&=& 
\delta(p-p') \left\{\delta(\Mybf{x}-\Mybf{x'})+ \cdots \right\}
\frac{\mu^{-1}(\Mybf{x})}{(N-1)!}\,,
\label{U-ort}
\ea
and calculate the integration measure $\mu(\Mybf{x})$. Here
$\delta(\Mybf{x}-\Mybf{x'})\equiv \prod_{k=1}^{N-1}\delta(x_k-x'_k)$ and
${\Mybf{x}}=(x_1,\ldots,x_{N-1})$ take positive real values, $x_k > 0$. Ellipses
denote the terms with all possible permutations inside the set
$\Mybf{x}=(x_1,\ldots,x_{N-1})$.

The calculation of \re{U-ort} repeats similar analysis for the closed spin chain
described at length in Ref.~\cite{SoV}. Substitution of \re{B-ei} into \re{U-ort}
yields
\be
\vev{p',\Mybf{x'}|p,\Mybf{x}}=(pp')^{Ns-1/2} \int \mathcal{D} w_N
\e^{ip\,w_N}\int \mathcal{D} w_N'\e^{-ip'\,\bar w_N'}
\vev{w_N',\Mybf{x}'|w_N,\Mybf{x}}\,,
\label{limit}
\ee
where the notation was introduced for the ket-vector
$\vev{z_1,\ldots,z_N|w_N,\Mybf{x}}=U_{\mybf{x}}(\vec z\,;\bar w_N)$, or
equivalently
\be
\ket{w_N,\Mybf{x}}=\Lambda_N(x_1)\Lambda_{N-1}(x_2)\ldots
\Lambda_2(x_{N-1})\ket{w_N}\,,
\label{L-comp}
\ee
with $\ket{w_N}$ being a ``single-particle state''. We recall that the operator
$\Lambda_k(u)$ maps $(k-1)-$particle state into $k-$particle one, so that a
composition of the $\Lambda-$operators in \re{L-comp} produces the $N-$particle
state. Calculating the scalar product $\vev{w_N',\Mybf{x}'|w_N,\Mybf{x}}$ one
applies systematically the exchange relation \re{exchange} and obtains
\be
\vev{w_N',\Mybf{x}'|w_N,\Mybf{x}}~=~c(\Mybf{x},\Mybf{x}')
\vev{w_N'|\lr{\Lambda_2^\dagger(x_{N-1}')\Lambda_2(x_1)} \ldots
\lr{\Lambda_2^\dagger(x_{1}')\Lambda_2(x_{N-1})}|w_N}\,,
\label{ww0}
\ee
where $c(\Mybf{x},\Mybf{x}')=\prod_{1\le j,\,k\le N-2 \atop j + k \le
N-1}\varphi(x_j,x_k')$. Notice that the exchange relation \re{exchange} holds
only for $x\neq y$. Therefore, calculating \re{ww0} we have tacitly assumed that
$x_j\neq x_k'$ for $j + k \le N-1$, or equivalently that all factors
$\varphi(x_j,x_k')$ are finite. For $N\ge 3$ the matrix element entering \re{ww0}
can be represented as follows
\be\label{ww}
\int \mathcal{D} w_1 \ldots \int \mathcal{D} w_{N-2}
\vev{w_N',x'_{N-1}|w_1,x_1}\vev{w_1,x'_{N-2}|w_2,x_2}\ldots
\vev{w_{N-2},x'_{1}|w_{N},x_{N-1}}\,,
\ee
where $\vev{w',x'|w,x}=[\Lambda_2^\dagger(x')\,\Lambda_2(x)](w';w)$. Thus, the
calculation of the scalar product \re{limit} for arbitrary $N$ is reduced to the
calculation of $\vev{w',x'|w,x}$ at $N=2$. Given that the separated variables at
$N=2$ take real positive values we deduce from \re{ww0} that the same holds true
for arbitrary $N$.

\begin{figure}[t]
\psfrag{=}[cc][cc]{$=$} \psfrag{a}[cc][cc]{$\beta_{x'}$}
\psfrag{b}[cl][cc]{$\alpha_{x'}$} \psfrag{d}[cc][cc]{$\beta_{x}$}
\psfrag{c}[cl][cc]{$\alpha_{x}$} \psfrag{a1}[cr][cc]{$\beta_{x'}+\epsilon$}
\psfrag{d1}[cr][cc]{$\beta_{x}+\epsilon$}

\psfrag{w}[cc][cc]{$w$}\psfrag{w1}[cc][cc]{$w'$}

\centerline{\epsfysize7.0cm\epsfbox{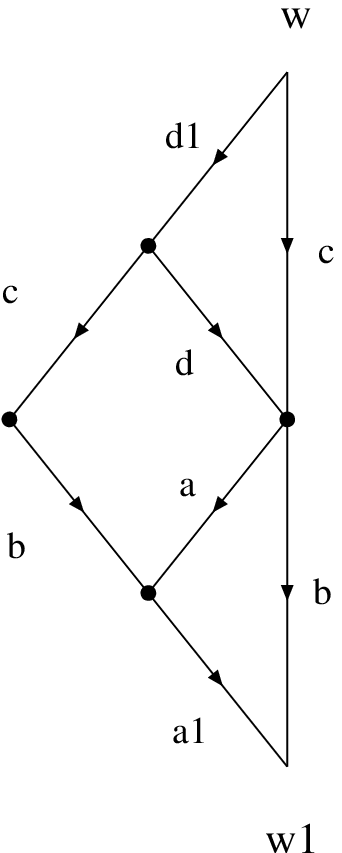}} \vspace*{0.5cm} \caption[]{The
scalar product $\vev{w',x'|w,x}$ for $N=2$}
\label{osp}
\end{figure}

To calculate the scalar product at $N=2$ we apply the diagrammatical approach of
Ref.~\cite{SoV} and represent the matrix element $\vev{w',x'|w,x}$ as the Feynman
diagram shown in Fig.~\ref{osp}. One expects from \re{U-ort} that
$\vev{w',x'|w,x}\sim \delta(x-x')$, so that the scalar product $\vev{w',x'|w,x}$
should be understood as a distribution. To find its explicit form we regularize
the corresponding Feynman integral by introducing a small parameter $\epsilon$
and shifting the indices of two lines as indicated in Fig.~\ref{osp}. Under such
regularization, the Feynman integral remains finite at $x=x'$ and it can be
calculated exactly with a help of the chain relation and permutation identity
(see Figs.~\ref{Chain} and \ref{comm-f}). The calculation is straightforward and
some details can be found in Ref.~\cite{SoV}. Going over to the momentum
representation and taking the limit $\epsilon\to 0$, one finds (for $x\,,x' >0$)
\be
\int \mathcal{D} w\,\e^{ipw}\, \vev{w',x'|w,x}~ =~2\pi\,\e^{ipw'}\,
p^{-2s}\,\Gamma^5(2s)\,\frac{|\Gamma(i(x+x'))|^2}{|\Gamma(s+ix)\Gamma(s+ix')|^4}
\,\delta(x-x')\,.
\label{Fourier-2}
\ee
Substituting \re{ww0} and \re{ww} into \re{limit} and taking into account
\re{Fourier-2} one obtains after some algebra
\ba
\vev{p',\Mybf{x'}|p,\Mybf{x}}&=&(2\pi)^{N-1} \delta(p-p')
\prod_{k=1}^{N-1}\delta(x_k-x_{N-k}')\cdot\Gamma^{N}(2s)\,\prod_{k=1}^{N-1}
\left[\frac{\Gamma(s-ix_k)\Gamma(s+ix_k)}{\Gamma(2s)}\right]^{-2N}
\nonumber\\[2mm]
& & \hspace*{-20mm}\times\left(\prod_{1\leq j\leq k\leq
  N-1}\frac{x_k+x_j}{\pi}\sinh\pi(x_k+x_j) \prod_{1\leq j<k\leq
  N-1}\frac{x_k-x_j}{\pi}\sinh\pi(x_k-x_j)\right)^{-1}\,.
\label{measure-1}
\ea
We recall that the calculation was performed under assumption that $x_j\neq x_k'$
for $j + k \le N-1$. Since the kernel $U_{p,\mybf{x}}$ is a symmetric function of
$\Mybf{x}$, $\vev{p',\Mybf{x'}|p,\Mybf{x}}$ should possess the same property.
This allows one to relax the above assumption and replace the product of
delta-functions $\prod_{k=1}^{N-1}\delta(x_k-x_{N-k}')$ in the r.h.s.\ of
\re{measure-1} by the sum $\sum_{\cal S}\delta(\Mybf{x}-{\cal S}\Mybf{x}')$ over
all permutations inside the set $\Mybf{x}'=\lr{x'_1,\ldots,x'_{N-1}}$.

Matching \re{measure-1} into \re{U-ort}, one finds the expression for the
integration measure in the SoV representation
\ba\label{mu}
\mu(\Mybf{x})&=&\frac{\Gamma^{-N}(2s)}{(N-1)!(2\pi)^{N-1}}
\prod_{k=1}^{N-1}\left[\frac{\Gamma(s-ix_k)\Gamma(s+ix_k)}{\Gamma(2s)}\right]^{2N}
\\[3mm]
&&\times\prod_{1\leq j<k\leq
  N-1}\frac{x_k^2-x_j^2}{2\pi^2}\left[\cosh(2\pi x_k)-\cosh(2\pi x_j)\right]\prod_{k=1}^{
  N-1}\frac{2x_k}{\pi}\sinh(2\pi x_k)\,.
\nonumber
\ea
This expression has the following properties. As expected, $\mu(\Mybf{x})$ is an
even function of the separated variables. It takes nonnegative values for real
$\Mybf{x}=(x_1,\ldots,x_{N-1})$ and vanishes on the hyperplanes $x_j=x_k$.

After analytical continuation to complex $\Mybf{x}$, the measure
$\mu({\Mybf{x}})$ becomes a meromorphic function of $x_k$ $(k=1,\ldots,N-1$) with
poles of the order $2N$ located along the imaginary axis at $x_k=\pm i(s+n)$ with
$n\in\mathbb{N}$. The measure decreases exponentially fast when one of the
separated variables, say $x_k$, goes to infinity along the real axis
\be\label{m-lim}
\mu(\Mybf{x}) ~\sim \,\e^{-2\pi |x_k|}x_k^{2Ns-3}\,.
\ee
One verifies that the measure \re{mu} satisfies the functional relation
\be
\frac{\mu(\Mybf{x}+i\Mybf{e}_k)}{\mu(\Mybf{x})}=\frac{x_k+i}{x_k}
\lr{\frac{x_k+is}{x_k+i(1-s)}}^{2N}\prod_{j\neq
k}\frac{x_k-x_j+i}{x_k-x_j}\frac{x_k+x_j+i}{x_k+x_j}\,,
\label{measure-rec}
\ee
with $\Mybf{e}_k$ defined in \re{Bax-U}.


It is instructive to compare \re{mu} with a similar expression for the
integration measure for the \textit{closed} spin chain \cite{SoV}
\be\label{mfi}
\mu_{\rm cl}(\Mybf{x})=\prod_{j,k=1\atop
j<k}^{N-1}{(x_k-x_j)}\,\sinh(\pi(x_k-x_j)) \,
\prod_{k=1}^{N-1}\left[\Gamma(s+ix_k)\Gamma(s-ix_k)\right]^N\,.
\ee
One observes that $\mu_{\rm cl}(\Mybf{x})$ enters as a factor into the expression
for $\mu(\Mybf{x})$, Eq.~\re{mu}.

\subsection{Eigenfunctions in the SoV representation}

The eigenfunctions $\Psi_{\mybf{q},p}(z_1,\ldots,z_N)$ are orthogonal to each
other for different sets of quantum numbers with respect to the $SL(2)$ scalar
product \re{norm}.  In the SoV representation the same condition looks as follows
\be
\vev{\Psi_{\mybf{q'},p'}|\Psi_{\mybf{q},p}}
= \vev{\Phi_{\mybf{q'}}|\Phi_{\mybf{q}}}_{_{\rm SoV}}\delta(p-p')
=\delta(p-p')\,\delta_{\mybf{q},\mybf{q'}}\,,
\label{ortho}
\ee
where the scalar product in the SoV representation is given by
\be
\vev{\Phi_{\mybf{q'}}|\Phi_{\mybf{q}}}_{_{\rm SoV}}= \int_{\mathbb{R}^+_{N-1}}
d^{N-1} \Mybf{x}\,\mu(\Mybf{x}) \lr{\Phi_{\mybf{q}'}(x_1,\ldots,x_{N-1})}^*
\Phi_{\mybf{q}}(x_1,\ldots,x_{N-1})\,.
\label{SoV-sp}
\ee
We recall that the momentum $p$ takes real positive values whereas the spectrum
of the integrals of motion $\Mybf{q}=(q_2,\ldots,q_N)$ is discrete~\cite{DKM99}.

To define the eigenfunction in the SoV representation,
$\Phi_{\mybf{q}}(\Mybf{x})$, we substitute $\Psi_{\mybf{q},p}(z_1,\ldots,z_N)$ in
\re{eit} by its integral representation \re{SoV-gen}. Following the standard
procedure~\cite{Sklyanin} and making use of Eqs.~\re{Bax-U} and \re{measure-rec} one can
show that $\Phi_{\mybf{q}}(\Mybf{x})$ satisfies the $(N-1)-$dimensional Baxter
equation
\be
\label{PhiB}
t_{N}(x_k) \,\Phi_{\mybf{q}}(\Mybf{x})~=~ \Delta_+(x_k)
\Phi_{\mybf{q}}(\Mybf{x}+i\Mybf{e}_k)~+~ \Delta_-(x_k)
\Phi_{\mybf{q}}(\Mybf{x}-i\Mybf{e}_k)\,,
\ee
where $t_N(x_k)$ is the eigenvalue of the transfer matrix, Eq.~\re{eit}.
As before, to solve this equation one has to specify additional conditions for
$\Phi_{\mybf{q}}(\Mybf{x})$.

Using \re{SoV-inv} one can show that $\Phi_{\mybf{q}}(\Mybf{x})$ is a polynomial
in $\Mybf{x}=(x_1,\ldots,x_{N-1})$. The proof goes along the same lines as
analysis of analytical properties of the Baxter $\mathbb{Q}-$operator in
Sect.~3.3. Namely, substituting the expression for the kernel \re{B-ei} into
\re{SoV-inv} and applying \re{drift}, one can express the r.h.s.\ of \re{SoV-inv}
as a nested contour integral. Analytical properties of the function
$\Phi_{\mybf{q}}(\Mybf{x})$ are in the one-to-one correspondence with the
properties of this integral.

It is easy to see that polynomial solutions to \re{PhiB} can be represented in
the factorized form
\be\label{PhiQ}
\Phi_{\mybf{q}}(\Mybf{x})~=~c_{\mybf{q}}\, Q_{\mybf{q}}(x_1)\ldots
Q_{\mybf{q}}(x_{N-1})\,,
\ee
where $Q_{\mybf{q}}(x)$ is the eigenvalue of the Baxter $\mathbb{Q}-$operator,
Eq.~\re{roots}, and the coefficient $c_{\mybf{q}}$ is fixed by the normalization
condition \re{ortho}. Substituting \re{PhiQ} into \re{SoV-sp} and taking into
account that $Q_{\mybf{q}}(x)$ is a real function, Eq.~\re{roots}, we find that
the solutions to the Baxter equation satisfy the orthogonality condition
\be\label{Q-ort}
\int_{\mathbb{R}^{N-1}_+} d^{N-1}\Mybf{x}\,\mu(\Mybf{x})
\prod_{k=1}^{N-1}Q_{\mybf{q}'}(x_k) \, Q_{\mybf{q}}(x_k) ~\sim
\delta_{\mybf{q},\mybf{q}'}\,,
\ee
with the measure given by \re{mu}.

Thus, having determined the eigenvalues of Baxter operator $Q_{\mybf{q}}(x)$ one
would be able to restore both the energy and the corresponding eigenfunction,
Eqs.~\re{QE} and \re{SoV-gen}, respectively. The solutions to the Baxter equation
for the $SL(2,\mathbb{R})$ open spin chain have been studied in \cite{DKM99}. It
turns out that the ground state $\Omega_p(\vec z)$ of the model can be found
exactly. This state has the total $SL(2)$ spin $h=0$,
Eq.~\re{S2-eig}, and has the form
~\footnote{In the terminology of the Algebraic Bethe Ansatz, $\Omega_p(\vec z)$
is a pseudovacuum state.}
\be
\Omega_p(\vec z)=\frac{p^{2s-1}}{\Gamma(2s)}\int \mathcal{D} w\, \e^{ipw}
\prod_{k=1}^N (z_k-\bar w)^{-2s}\,.
\label{pseudo}
\ee
Indeed, $\Omega_p(\vec z)$ diagonalizes simultaneously the operators of
two-particle spins, $(J_{n,n+1}-2s)\Omega_p(\vec z)=0$, Eq.~\re{JJ}, and leads
the energy $E_{\mybf{q}}=0$. It is interesting to notice that $\Omega_p(\vec z)$
is also the ground state of the homogeneous $SL(2,\mathbb{R})$  \textit{closed}
spin chain~\cite{SoV}.

The state \re{pseudo} diagonalizes the Baxter operator
$\mathbb{Q}(u)\ket{\Omega_p} = \ket{\Omega_p}$, or equivalently
$Q_{\mybf{q}}(u)=1$. As a consequence, it admits the following integral
representation (see Ref.~\cite{SoV})
\footnote{Notice that this relation takes the same form both for the open and
closed spins chain whereas the expressions for the integration measure and the
transition kernel to the SoV representation are different in the two cases.}
\be
\Omega_p(\vec z)=p^{Ns-1/2}\e^{-i\pi s(2N-1)}\int_{\mathbb{R}^{N-1}_+}
d^{N-1}\Mybf{x}\,\mu(\Mybf{x}) U_{p,\mybf{x}}(\vec z)\,.
\label{Omega-SoV}
\ee
Let us calculate the $SL(2)$ scalar product $\vev{\Omega_p|\Omega_{p'}}$ and use
two different expressions for $\Omega_p(\vec z)$, Eqs.~\re{pseudo} and
\re{Omega-SoV}. Equating the two expressions, one finds that the integration
measure \re{mu} satisfies the normalization condition
\be\label{mN}
\int_{\mathbb{R}^{N-1}_+} d\Mybf{x}\,\mu(\Mybf{x}) ~=~\frac{1}{\Gamma(2Ns)}\,.
\ee

The relation \re{Omega-SoV} allows one to establish the equivalence between the
SoV and the ABA methods for the $SL(2,\mathbb{R})$ open spin chain. Let
$\lambda_1,\ldots,\lambda_h$ be the Bethe roots, or equivalently zeros of the
polynomial $Q_{\mybf{q}}(x)$, Eq.~\re{roots}. Applying $\widehat
B(\lambda_1)\ldots \widehat B(\lambda_h)$ to the both sides of \re{Omega-SoV} and
taking into account \re{B-pol} we obtain
\be\label{ABASOV}
\Psi_{p,\mybf{q}}(\vec z)=\widehat B(\lambda_1)\ldots \widehat B(\lambda_h) \,
\Omega_p(\vec{z})~=~ c(p)\int_{\mathbb{R}^{N-1}_+} d\Mybf{x}\,\mu(\Mybf{x})\,
U_{p,x}(\vec{z})\prod_{k=1}^{N-1}Q_{\mybf{q}}(x_k)\,,
\ee
where $c(p)$ is the normalization factor. In \re{ABASOV}, the first relation
coincides with the ABA representation for the eigenstate of the model while the
second one defines the same eigenstate in the SoV representation.

The explicit form of the eigenfunctions in the SoV representation \re{PhiQ}
suggests that there exists a relation between the Baxter $\mathbb{Q}-$operator
and the transition kernel $U_{p,\mybf{x}}(\vec{z})$~\cite{KuS}. In the case of
the closed spin chain it has been established in Ref.~\cite{SoV}. It turns out
that this relation is universal and it also holds for the open spin chain
\be\label{QU}
 U_{p,\mybf{x}}(\vec{z})~=~[\mathbb{Q}(x_1)\,\ldots\mathbb{Q}(x_{N-1})\,
\Theta_p\,](z_1,\ldots,z_N)\,.
\ee
Here 
$\Theta_p(\vec z)$ is a certain $\Mybf{x}-$independent function of $\vec z$,
which does not belong to the quantum space of the model.%
\footnote{In Ref.~\cite{SoV}, the function $\Theta_p(\vec z)$ was defined as a
limiting case of the state $\ket{\Omega_{\bar w_0,\bar w_N}}$ belonging to the
Hilbert space of the model.} Since $\mathbb{Q}(is)=\mathbb{K}$, the function
$\Theta_p(\vec z)$ is equal to $U_{p,\mybf{x}}(\vec{z})$ for special values of
the $x-$variables, $x_1=\ldots=x_{N-1}=is$ that we denote as
$U_{p,{is}}(\vec{z})$. The expression for $U_{p,{is}}(\vec{z})$ can be easily
obtained from diagrammatic representation of the kernel (see Fig.~\ref{opyr})
\be\label{exOm}
\Theta_p(\vec{z})~=~U_{p,{is}}(\vec{z})~=~p^{Ns-1/2}\e^{i\pi s(N-1)}\,
\e^{ipz_{N}}\,.
\ee
Notice that $\Theta_p(\vec{z})$ depends only on a single variable $z_N$ and,
therefore, it is not normalizable with respect to the $SL(2)$ scalar product
\re{norm}.

The proof of \re{QU} can be performed diagrammatically and it repeats similar
analysis in Ref.~\cite{SoV}. Another way to verify \re{QU} is to use the
following identities
\ba
&&\hspace*{-10mm}\int\mathcal{D}^N\!w \, Q_u(\vec z|\vec w)\,\e^{ipw_{N}} =
\e^{-i\pi s}\int \mathcal{D} w_N\, \Lambda_u(z_{N-1},z_N|\bar w_N) \e^{ipw_{N}}
\label{Q-red}
\\
&&\hspace*{-10mm}\int\mathcal{D}^N\!w \, Q_u(\vec z|\vec
w)\Lambda_v(w_k,\ldots,w_N|\bar y_{k+1},\ldots,\bar y_N) = \e^{-i\pi
s}[\Lambda_{k+1}(u)\Lambda_k(v)](z_{k-1},\ldots,z_N;\bar y_{k+1},\ldots,\bar
y_N)\,, \nonumber
\ea
with $\vec z=(z_1,\ldots,z_N)$, $\vec w=(\bar w_1,\ldots,\bar w_N)$ and
$\mathcal{D}^N w=\prod_{n=1}^N\mathcal{D} w_n$. We recall that the
$\Lambda-$operator was defined in Eqs.~\re{Lambda1} and \re{Lambda-ker}. To
derive \re{Q-red} one substitutes the expression for $Q_u(\vec z|\vec w)$,
Eq.~\re{Q-explicit}, and integrates over ``free'' vertices $w_1,\ldots,w_{k-1}$
with a help of the identity \re{drift} for $\Psi(w)=1$. Eqs.~\re{Q-red} can be
rewritten symbolically as
\be
\mathbb{Q}(u)\ket{\Theta_p} =  \e^{-i\pi s}\Lambda_2(u)\ket{\Theta_p} \,,\qquad
\mathbb{Q}(u)\Lambda_k(v)=\e^{-i\pi s}\Lambda_{k+1}(u)\Lambda_k(v)\,.
\ee
Applying these relations one verifies that \re{QU} coincides with \re{B-ei} and
\re{U}.

\section{Relation to the Wilson polynomials}

In this section, we consider the open spin chain with $N=2$ sites. We will
demonstrate that in that case the eigenvalues of the Baxter $\mathbb{Q}-$operator
are given by the Wilson polynomials~\cite{Koekoek} and the unitary transformation
to the SoV representation coincides with the Fourier-Jacobi
transformation~\cite{Koor}.

At $N=2$ the Hamiltonian of the open spin chain \re{H} equals
$\mathcal{H}_2=H_{12}=2[\psi(J_{12})-\psi(2s)]$. Its eigenstates are uniquely
fixed by the values of the momentum $p$ and the total $SL(2)$ spin $h$,
Eq.~\re{S2-eig} and are given by
\be
\Psi_{p,h}(z_1,z_2) = \frac{p^{2s-1}}{\Gamma(2s)}\int \mathcal{D} w\, \e^{ipw}
\frac{(z_1-z_2)^h}{(z_1-\bar w)^{2s+h}(z_1-\bar w)^{2s+h}}\,.
\label{Psi-N=2}
\ee
Substituting \re{Psi-N=2} into \re{Q-cont}, one finds the
eigenvalues of the $N=2$ Baxter operator after some algebra as%
\footnote{As was explained in Sect.~3.3, due to the $SL(2)$ invariance of the
Baxter operator, one can calculate $Q_h(u)$ by substituting
$\Psi(z_1,z_2)=(z_1-z_2)^h$ into \re{Q-cont}.}
\be
Q_h(u)={}_4 F_3\lr{{{-N,N+4s-1,s+iu,s-iu}\atop {2s,2s,2s}}\bigg|1}=\left[
\frac{\Gamma(2s)}{\Gamma(2s+N)}\right]^3W_N(u^2,s,s,s,s)\,,
\ee
where $W_N(u^2)$ is the Wilson polynomial~\cite{Koekoek}. It is interesting to
note that the solution of the $N=2$ Baxter equation for the closed spin chain is
given by the continuous Hahn polynomials (see e.g. Ref.~\cite{SoV}). The
polynomials $Q_h(x)$ are orthogonal on the half-axis $x>0$ with respect to the
scalar product \re{Q-ort} with the measure \re{mu} given by
\be
\mu_{N=2}(x) 
=\frac1{2\pi}\left|\frac{\Gamma^4(s+ix)}{\Gamma(2ix)\Gamma^3(2s)} \right|^2\,.
\ee
This property is in a perfect agreement with the orthogonality condition for the
Wilson polynomials \cite{Koekoek}.

The  eigenvalue of the $N=2$ Hamiltonian corresponding to \re{Psi-N=2} can be
calculated either by replacing the operator $J_{12}$ in the expression for
$\mathcal{H}_2$ by its eigenvalue $(J_{12}-h-2s)\Psi_{p,h}=0$, or by applying
\re{QE}. In this way, one obtains
\be\label{E-2}
E_h = \pm i\frac{d}{d\epsilon}\ln Q_{h}(\pm
is+\epsilon)\biggl|_{\epsilon=0}=2\left[\psi(h+2s)-\psi(2s)\right]\,.
\ee
The ground state corresponds to $h=0$.

Let us examine the unitary transformation to the SoV representation \re{SoV-gen}
for $N=2$. It is defined by the transition kernel, Eqs.~\re{B-ei} and \re{U},
which looks at $N=2$ like
\be
\label{U-N=2}
U_{p,x}(z_1,z_2)= p^{2s-1/2} \int \mathcal{D} w_2\, \e^{ipw_2}
\Lambda_x(z_1,z_2|\bar w_2)\,,
\ee
where $\Lambda_x(z_1,z_2|\bar w_2)$ are given by \re{Lambda-ker}
\be
\Lambda_x(z_1,z_2|\bar w_2)=\e^{3i\pi s}\int \mathcal{D} y_2 \,(y_2-\bar
w_2)^{-\beta_x}(z_2-\bar w_2)^{-\alpha_x}(z_1-\bar y_2)^{-\alpha_x}(z_2-\bar
y_2)^{-\beta_x}\,, \nonumber
\ee
with $\beta_x=s+ix$ and $\alpha_x=s-ix$. It is convenient to transform
$U_{p,x}(z_1,z_2)$ to the momentum representation.

For arbitrary function $\Psi(z_1,z_2)\in \mathcal{V}_2$ this transformation is
defined as
\be\label{F}
\Psi(z_1,z_2)~=~\frac{1}{\Gamma(2s)}\int_0^\infty dp_1\,dp_2\,
\e^{i(p_1z_1+p_2z_2)} (p_1\,p_2)^{s-1/2}\,\widetilde \Psi(p_1,p_2)\,,
\ee
where the additional factor $(p_1\,p_2)^{s-1/2}$ was introduced to simplify the
expression for the scalar product \re{norm} in the momentum representation (see
Eq.~\re{delta-f})
\be
\vev{\Psi_1 |\Psi_2} = \int_0^\infty dp_1 dp_2\,
\lr{\widetilde\Psi_1(p_1,p_2)}^*\widetilde\Psi_2(p_1,p_2)\,.
\ee
In particular, the $N=2$ eigenstates \re{Psi-N=2} are given in the momentum
representation by the Jacobi polynomials 
\footnote{This expression is well-known in QCD as defining the conformal
operators built from two fields with the conformal spin $s$.}
\be
\widetilde \Psi_{p,h}(p_1,p_2) = a_h\, \delta(p-p_1-p_2)
(p_1p_2)^{s-1/2}(p_1+p_2)^h\,
\textrm{P}^{(2s-1,2s-1)}_h\lr{\frac{p_1-p_2}{p_1+p_2}}\,,
\label{mom-eig}
\ee
where $a_h=i^{-h-4s} h!\Gamma(2s)/\Gamma^2(h+2s)$ and delta-function ensures the
momentum conservation. Applying \re{alpha-rep} and performing integration in
\re{U-N=2}, one finds that in the momentum representation the $N=2$ transition
kernel is given by
\be\label{UP}
\widetilde U_{p,x}(p_1,p_2)~=\delta(p-p_1-p_2)\,\frac{\Gamma^2(2s)\e^{i\pi
s}}{|\Gamma(s+ix)|^2} \left(\frac{p}{p_1p_2}\right)^{1/2}
\left(\frac{p_2}{p_1}\right)^s {}_2F_1\lr{{{s-ix,s+ix} \atop
{2s}}\bigg|-\frac{p_2}{p_1}}\,.
\ee
It defines the SoV transformation $\widetilde\Psi_p(p_1,p_2) \mapsto \Phi(x)$
\be\label{FU}
\Phi(x)\delta(p-p')~=\int_0^\infty dp_1\,dp_2 \left (\widetilde
U_{p',x}(p_1,p_2)\right)^*\,\widetilde \Psi_{p}(p_1,p_2)\,,
\ee
with $p\,,x>0$. Substituting \re{UP} into this relation and introducing notations
for $\widetilde \Psi_p(p_1,p_2)=\delta(p-p_1-p_2)
p^{-1/2}\xi^{s-1/2}(1+\xi)f(\xi)$ with $\xi=p_2/p_1$, one finds that \re{FU} is
reduced to
\be\label{FJ}
\Phi(x)~=~ \e^{i\pi s}\, \frac{\Gamma^2(2s)}{|\Gamma(s+ix)|^2}\int_0^\infty
d\xi\, \xi^{2s-1}\,{}_2F_1\lr{{{s-ix,s+ix} \atop {2s}}\bigg|-\xi}\, f(\xi)\,.
\ee
This relation defines the map $f(\xi)\mapsto \Phi(x)$, which is known as the
Fourier-Jacobi or the index hypergeometric transform~\cite{Koor}. Then, the
unitarity of the SoV transformation at $N=2$ follows from the similar property of
the transformation \re{FJ}.

Let us replace $\widetilde \Psi_p(p_1,p_2)$ in \re{FU} by the $N=2$ eigenstate
\re{mom-eig}. According to \re{PhiQ} and \re{SoV-inv}, the corresponding
eigenfunction in the separated variables is given by the Wilson polynomial
$\Phi(x)= Q_h(x)\sim W_h(x^2,s,s,s,s)$. Then, \re{FJ} leads to a known
representation for the Wilson polynomials as the index hypergeometric transform
of the Jacobi polynomials~\cite{Koor}.

\setcounter{equation}{0}
\section{Concluding remarks}

In this paper we have constructed the Baxter $\mathbb{Q}-$operator and the
representation of the Separated Variables for the open homogeneous
$SL(2,\mathbb{R})$ spin magnet. Our analysis relied on the diagrammatical
approach developed in Refs.~\cite{SoV} in application to the closed spin chain.
In this approach, one represents the kernels of the relevant integral operators
($\mathbb{Q}-$operator, the transition kernel to the SoV representation) as
Feynman diagrams and establishes their various properties with a help of a few
simple diagrammatical identities.

We found that the Feynman diagrams for the $\mathbb{Q}-$operator and the
transition kernel to the SoV representation have a remarkably simple form (see
Figs.~\ref{fig1} and \ref{opyr}). In the latter case, the diagram reveals a
universal pyramid-like structure which has been already observed for various
quantum integrable models like periodic Toda chain~\cite{KL}, closed
$SL(2,\mathbb{R})$ and $SL(2,\mathbb{C})$ spin chains~\cite{DKM,SoV} and
Calogero-Sutherland model~\cite{KMS}. This structure is a manifestation of a
general factorization property \re{U} of the transition kernel to the SoV
representation. Namely, the kernel is factorized into the product of
$\Lambda-$operators each depending on a single separated variable. The only
difference between the models mentioned above resides in the explicit form of the
$\Lambda-$operator. The latter can be obtained as a certain limit of the
$\mathbb{Q}-$operator leading to the expression for the transition kernel to the
SoV representation as the product of the $\mathbb{Q}-$operators projected onto a
special reference state. Another advantage of the diagrammatical approach is that
it offers a simple regular way for calculating the integration measure in the SoV
representation.

We found that there exists an intrinsic relation between the open spin chains and
Wilson polynomials~\cite{Koekoek}. The latter occupy the top level in the Askey
scheme of hypergeometric orthogonal polynomials~\cite{Askey}. These polynomials
define the eigenvalues of the Baxter operator for open spin chain with $N=2$
sites~\cite{DKM99}.

It is straightforward to extend our analysis to the case of inhomogeneous open
$SL(2,\mathbb{R})$ spin chains. One can show that for such models the transition
kernel to the SoV representation is given by the same pyramid-like diagram shown
in Fig.~\ref{opyr} with the only difference that both the indices attached to
various lines and the integration measure corresponding to internal vertices
should be modified appropriately. Another interesting possibility could be to
consider an open spin chain with the $SL(2,\mathbb{C})$ symmetry. Such models
naturally appear in high-energy QCD as describing Regge singularities of
scattering amplitudes with meson quantum numbers~\cite{KK}. In that case, the
quantum space of the model does not possess the highest weight (pseudovacuum
state) and, therefore, the Algebraic Bethe Ansatz is not applicable.

\section*{Acknowledgements}

We are grateful to V.~Braun and F.~Smirnov for useful discussions. This work was
supported in part by the grant 03-01-00837 of the Russian Foundation for
Fundamental Research, by the NATO Fellowship (A.M. and S.D.) and the Sofya
Kovalevskaya programme of Alexander von Humboldt Foundation (A.M.).

\appendix
\renewcommand{\theequation}{\Alph{section}.\arabic{equation}}
\setcounter{table}{0}
\renewcommand{\thetable}{\Alph{table}}

\section{Appendix: Feynman diagram technique}
\label{Ap}

Here we collect some useful formulae for the $SL(2,\mathbb{R})$ integrals. Their
derivation can be found in Refs.~\cite{SoV}.

\noindent%
\begin{figure}[ht]%
\vspace*{-5mm}
\psfrag{a}[cc][cc]{$\alpha$}
\psfrag{w}[cc][cc]{$\bar w$}
\psfrag{z}[cc][cc]{$z$} $\bullet$~Propagator:\vspace*{-2mm}
\be
{\parbox[c]{30mm}{\centerline{\epsfxsize20mm\epsfbox{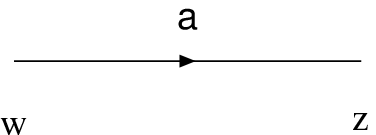}}}}
=\frac{1}{(z-\bar w)^\alpha}=\frac{\e^{-i\pi
\alpha/2}}{\Gamma(\alpha)}\,\int_0^{\infty} dp\,\e^{ip\,(z-\bar w)} \,
p^{\alpha-1}\,,
\label{alpha-rep}
\ee
\vspace*{-5mm}
\end{figure}%

\begin{figure}[t]
\psfrag{a}[cc][cc]{$\beta$} \psfrag{b}[cc][cc]{$\alpha$}
\psfrag{ab}[cc][cc]{$\alpha+\beta-2s$} \psfrag{text}[cc][cc]{$=\ \ \ \ \ \ \
a(\alpha,\beta)\ \ \ \ \times $} \centerline{\epsfxsize14.0cm\epsfbox{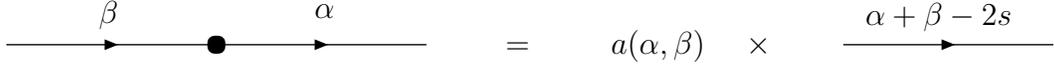}}
\vspace*{0.5cm} \caption[]{Chain relation.}
\label{Chain}
\end{figure}

\begin{figure}[ht]
\psfrag{a1}[cl][cc]{$s+iy$} \psfrag{a2}[cl][cc]{$s-iy$}
\psfrag{a3}[cl][cc]{$s-ix$} \psfrag{a4}[cl][cc]{$s+ix$}
\psfrag{a5}[bc][cc][1][90]{$i(y-x)$} \psfrag{b1}[cr][cc]{$s+iy$}
\psfrag{b2}[cr][cc]{$s-iy$} \psfrag{b3}[cr][cc]{$s-ix$}
\psfrag{b4}[cr][cc]{$s+ix$} \centerline{\epsfxsize10.0cm\epsfbox{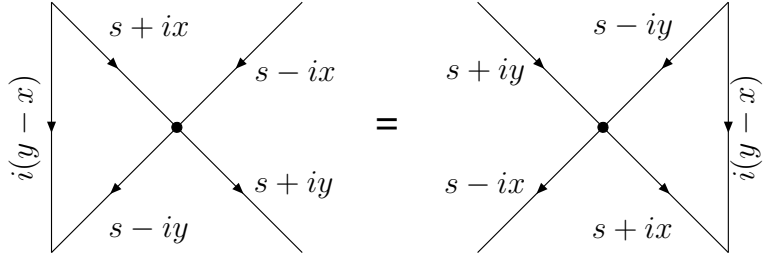}}
\vspace*{0.5cm} \caption[]{Permutation identity.}
\label{comm-f}
\end{figure}

\begin{figure}[h!]
\psfrag{a1}[cl][cc]{$s+iy$} \psfrag{a2}[cl][cc]{$s-iy$}
\psfrag{a3}[cl][cc]{$s-ix$} \psfrag{a4}[cl][cc]{$s+ix$}
\psfrag{a5}[bc][cc][1][90]{$i(y-x)$} \psfrag{b1}[cr][cc]{$s+iy$}
\psfrag{b2}[cr][cc]{$s-iy$} \psfrag{b3}[cr][cc]{$s-ix$}
\psfrag{b4}[cr][cc]{$s+ix$} \centerline{\epsfxsize10.0cm\epsfbox{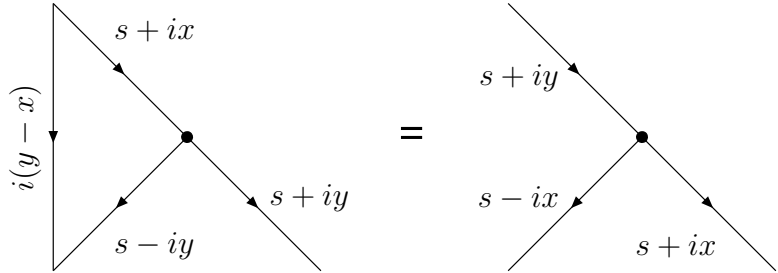}}
\vspace*{0.5cm} \caption[]{Special case of the permutation identity. It is
obtained {}from Figure~\ref{comm-f} by sending one of the external points to
infinity.}
\label{amp1}
\end{figure}

\medskip
\noindent$\bullet$~``Chain relation'' (see Fig.~\ref{Chain}):
\be
\int \mathcal{D}w\,{(z-\bar w)^{-\alpha} (w-\bar v)^{-\beta}~=~
{a(\alpha,\beta)}\,(z-\bar v)^{-\alpha-\beta+2s}}\,,\qquad (\alpha+\beta\neq 2s)
\label{chain-h}
\ee
$\bullet$~Delta-function relation:
\be
\int \mathcal{D} w  \e^{i p w-ip'\bar w} = \delta(p-p')
\,p^{1-2s}\cdot\Gamma(2s)\,,
\label{delta-f}
\ee
$\bullet$~Identity operator:
\be
[\mathbb{K}\cdot\Psi](z) = \int \mathcal{D}w\,\frac{\e^{i\pi s}}{(z-\bar
w)^{2s}}\Psi(w)= \Psi(z)\,.
\label{K-ker}
\ee
$\bullet$~Contour-integral representation:
\be
\int\mathcal{D}w\,\frac{\e^{i\pi s}\Psi(w)}{(z_1-\bar w)^{\alpha_x} (z_2-\bar
w)^{\beta_x}}=\frac{\Gamma(2s)}{\Gamma(\alpha_x)\Gamma(\beta_x)} \int_0^1d\tau
\tau^{\alpha_x-1}(1-\tau)^{\beta_x-1}\Psi(\tau z_1+(1-\tau)z_2)\,.
\label{drift}
\ee
$\bullet$~Fourier integral:
\be
\int \mathcal{D} w\, \frac{\e^{i p w}}{(z-\bar{w})^{\alpha}} =
 \theta(p)\,p^{\alpha-2s}\e^{i p z}\cdot\e^{-i\pi\alpha/2}\frac{\Gamma(2s)}{\Gamma(\alpha)}\,,
\label{p-rep}
\ee
$\bullet$~Permutation identity (see Figs.~\ref{comm-f} and \ref{amp1}):
\be
\label{comm-i}
(z_2-\bar v_2)^{i(x-y)}I(z,\bar v\,;x,y)=(z_1-\bar v_1)^{i(x-y)} I(z,\bar
v\,;y,x)\,,
\ee
where $z=(z_1,z_2)$, $\bar v=(\bar v_1,\bar v_2)$ and
\be
I(z,\bar v\,;x,y)= \int \mathcal{D}w \frac{1}{(w-\bar v_1)^{\alpha_x}(w-\bar
v_2)^{\beta_x} (z_1-\bar w)^{\beta_y}(z_2-\bar w)^{\alpha_y}}\,.
\label{four}
\ee
In these relations, $\alpha_x=s-ix$ and $\beta_x=s+ix$ for arbitrary $x$, the
$a-$function is defined in \re{a} and the integration measure $\mathcal{D} w$ is
given by \re{measure}, $\bar w=w^*$ and $p>0$.

\begin{thebibliography}{99}

\bibitem{QISM} L.A.~Takhtajan and L.D.~Faddeev, Russ.\ Math.\ Survey\ {\bf 34} (1979) 11;
\\        E.K.~Sklyanin, L.A.~Takhtajan and L.D.~Faddeev,
          Theor.\ Math.\ Phys.\ {\bf 40} (1980) 688;
\\        V.E.~Korepin, N.M.~Bogoliubov and A.G.~Izergin, {\it Quantum
          inverse scattering method and correlation functions\/},
          Cambridge Univ. Press, 1993.
\bibitem{BDM}  V.M.~Braun, S.\'E.~Derkachov, A.N.~Manashov,
          Phys.\ Rev.\ Lett.\ \textbf{81} (1998) 2020.
\bibitem{AB}   A.V.~Belitsky,
          Phys.\ Lett.\ B \textbf{453} (1999) 59;
          Nucl.\ Phys.\ B \textbf{558} (1999) 259;
          Nucl.\ Phys.\ B \textbf{574} (2000) 407.
\bibitem{DKM99}
          S.~\'E.~Derkachov, G.~P.~Korchemsky and A.~N.~Manashov,
          Nucl.\ Phys.\ B {\bf 566} (2000) 203.
\bibitem{Baxter}
          R.J.~Baxter, {\it Exactly Solved Models in Statistical
          Mechanics\/}, Academic Press, London, 1982;
          Stud.\ Appl.\ Math.\ {\bf 50} (1971) 51.
\bibitem{ABA}  L.D.~Faddeev,
          Int.\ J.\ Mod.\ Phys.\ A {\bf 10} (1995) 1845. 
\bibitem{Sklyanin}
          E.K.~Sklyanin, {\it The quantum Toda chain\/},
          Lecture Notes in Physics, vol.\ 226, Springer, 1985, pp.196--233;
          {\it Functional Bethe ansatz\/}, in ``Integrable
          and superintegrable systems'', ed.\ B.A. Kupershmidt, World
          Scientific, 1990, pp.8--33;
          {\it Quantum inverse scattering method. Selected topics\/},
          ``Quantum Group and Quantum Integrable Systems'' (Nankai
          Lectures in Mathematical Physics), ed.\ Mo-Lin Ge, Singapore: World Scientific,
          1992, pp.\ 63--97 [hep-th/9211111];
          Progr.\ Theor.\ Phys.\ Suppl.\ {\bf 118} (1995) 35
          [solv-int/9504001].
\bibitem{PG} V.~Pasquier amd M.~Gaudin, J.~Phys.~A: Math.~Gen.
          {\bf 25} (1992) 5243.
\bibitem{KL}   S.~Kharchev and D.~Lebedev,
          Lett.\ Math.\ Phys.\  {\bf 50} (1999) 53;
          JETP Lett.\  {\bf 71} (2000) 235;
          J.\ Phys.\ A: Math. Gen. {\bf 34} (2001) 2247. 
\bibitem{KSS}
          V.B.~Kuznetsov, M.Salerno and E.K.~Sklyanin,
          J.\ Phys.\ A {\bf 33} (2000) 171.
\bibitem{SD}
          S.\'E. Derkachov, J. Phys. A: Math. Gen. {\bf 32} (1999) 5299.
\bibitem{DKM}S.\'E.~Derkachov, G.P.~Korchemsky and A.N.~Manashov,
          Nucl.\ Phys.\ B {\bf 617} (2001) 375. 
\bibitem{SoV} S.\'E.~Derkachov, G.~P.~Korchemsky and A.~N.~Manashov,
          JHEP {\bf 0307} (2003) 047.
\bibitem{KMS}
V.~B.~Kuznetsov, V.~V.~Mangazeev and E.~K.~Sklyanin,
arXiv:math.ca/0306242.

\bibitem{Sklyanin88}  E.K.~Sklyanin, J.Phys.A: Math.Gen. {\bf 21}
           (1988) 2375.
\bibitem{Gelfand}
          I.M.~Gelfand, M.I.~Graev and N.Ya.~Vilenkin, {\it Generalized functions. Vol.5
          Integral geometry and representation theory\/}, New York, NY Academic Press 1966.
\bibitem{Ch}
    I.V.~Cherednik,
    Theor.\ Math.\ Phys.\ {\bf 61} (1984) 977.

\bibitem{KS}
    P.P.~Kulish and E.K.~Sklyanin,
    J.\ Phys.\ A {\bf 25} (1992) 5963;
\\
    P.P.~Kulish and R.~Sasaki,
    Prog.\ Theor.\ Phys.\ {\bf 89} (1993) 741;
\\
    H.J.~de Vega and A.~Gonzalez-Ruiz,
    J.\ Phys.\ {\bf A27} (1994) 6129.
\bibitem{Koor}
          T.H.~Koornwinder, 
          Lect.~Notes Math. {\bf 1171} (1985) 174. 

\bibitem{KuS} V.B.~Kuznetsov and E.K.~Sklyanin,
          J.\ Phys.\ A {\bf 31} (1998) 2241. 

\bibitem{Koekoek} R.~Koekoek and S.~Swarttouw, {\it The Askey scheme of hypergeometric
orthogonal polynomials and its q-analogues}, Report 94-05, Delft University of
Technology, 1994. 

\bibitem{Askey}
G.E.~Andrews, R.~Askey and R.~Roy, \textit{Special functions}, Encyclopedia of
Mathematics and Its Applications, The University Press, Cambridge, 1999.

\bibitem{KK}
D.~Karakhanian and R.~Kirschner,
Phys.\ Atom.\ Nucl.\  {\bf 65} (2002) 1501 [Yad.\ Fiz.\  {\bf 65} (2002) 1539].



\end{thebibliography}
\end{document}